\documentstyle[seceq,mbf,epsf,wrapft]{ptptex}
\notypesetlogo

\newcommand{\qsp}{\stackrel{{\rm qs}}{\rightarrow}}
\newcommand{\tps}{\stackrel{{\rm a}}{\rightarrow}}
\newcommand{\fwdp}{{\rightarrow}_*}
\newcommand{\pderf}[3]{
\left( {\partial #1 \over \partial #2} \right)_{#3}}
\newcommand{\bra}{\left\langle}
\newcommand{\ket}{\right\rangle}
\newcommand{\pder}[2]{
{\partial #1 \over \partial #2} }
\newcommand{\pdert}[2]{
{\partial^2 #1 \over \partial #2^2} }
\newcommand{\der}[2]{
{d #1 \over d #2} }
\newcommand{\rev}{{\cal R}}

\title{
Thermodynamic Irreversibility from high-dimensional 
Hamiltonian Chaos
}
\author{
Shin-ichi {\sc Sasa}\footnote{
E-mail address: sasa@jiro.c.u-tokyo.ac.jp
} 
and Teruhisa S. {\sc Komatsu}\footnote{E-mail address: 
komatsu@jiro.c.u-tokyo.ac.jp }
}

\inst{
Department of Pure and Applied Sciences, University of Tokyo, \\
Komaba, Meguro-ku, Tokyo 153, Japan
}

\recdate{
November 9, 1999
}

\abst{
This paper discusses the thermodynamic irreversibility realized 
in high-dimensional Hamiltonian systems with a time-dependent 
parameter. 
A new quantity, the irreversible information loss, is defined  from 
the Lyapunov analysis so as to characterize the thermodynamic 
irreversibility. It is proved that this new quantity satisfies an 
inequality associated with the second law of thermodynamics.
Based on the assumption that these systems possess the mixing property
and certain large deviation properties in the thermodynamic limit, 
it is argued reasonably that the most probable value of the irreversible 
information loss is equal to the change of the Boltzmann entropy 
in statistical mechanics, and that it is always a non-negative value.
The consistency of our argument is confirmed by numerical experiments
with the aid of the definition of a quantity we refer to as the
excess information loss. 
}

\begin{document}

\maketitle

\section{Introduction}

Thermodynamics formalizes a fundamental limitation of possible 
processes between equilibrium states. In particular, when a thermodynamic 
system is enclosed by  adiabatic walls,  the limitation is represented by, 
for example, a fact that, given a system in some initial state, 
it is not possible to lower the system's energy by first changing some of 
its other extensive variables and then returning them to their original 
values. Contrastingly, the energy of the system can be increased
by the similar change of the other extensive variables.
These two facts make clear the special nature of energy as an extensive 
variable. This asymmetry is the basis of  {\it thermodynamic irreversibility}.

Thermodynamics is one of the most elegant theories being based on only 
a few fundamental principles. \cite{Lieb} However, one may wonder how 
its principles emerge out of purely mechanical systems.
Thermodynamic systems consist of many  molecules, whose dynamics are 
described by Hamiltonian equations. 
Thus, in the idealized limit of adiabatic walls, 
a thermodynamic system can be regarded as a Hamiltonian system that 
is connected to some mechanical apparatus,  but does not contact a heat 
reservoir.  With this in mind, it may be natural to  expect
that thermodynamic irreversibility can be formalized in Hamiltonian systems.   

Thermodynamic entropy plays a central role in the description
of  thermodynamic irreversibility, and the thermodynamic entropy is generally
thought to be given by the logarithm of the number of micro-states. This
relation, the Boltzmann formula, seems well-established as far as the 
calculation of equilibrium  values is concerned. 
However, it has not yet been shown that the Boltzmann formula provides 
a complete account of irreversibility.\cite{Lebo}

In this paper, we  discuss  thermodynamic irreversibility based 
on  the nature of high-dimensional Hamiltonian chaos. As our most 
notable result, we find a new quantity which satisfies an inequality 
associated with thermodynamic irreversibility. We define this quantity, 
the ``irreversible information loss'', from  dynamical system considerations.
Furthermore, we argue reasonably that 
the irreversible information loss is related to the change of the 
Boltzmann entropy, and this leads us to conclude that   the Boltzmann
entropy does not decrease for any processes in the thermodynamic limit.

\subsection{Related studies}

This paper provides theoretical arguments for numerical results 
reported in a previously published paper, \cite{SK} and contains 
a detailed description of the numerical experiment. 

The present study has been carried out under the influence of 
several related studies.  First, the  attempt to construct
steady state thermodynamics by Oono and Paniconi has 
provided the direction of the present study \cite{Oono}. 
They have proposed an operational method to obtain 
non-equilibrium thermodynamic functions. In addition,
from a more general viewpoint, they emphasize
the importance of theory concerning the relation between two different states.
Following this manner of thinking, we have set out to study thermodynamic 
irreversibility from dynamical systems. 

The stochastic energetics proposed by Sekimoto has given a nice example 
of the construction of thermodynamics from dynamical systems.\cite{Ken} 
Stochastic energetics formalizes  energy transformation in  Langevin 
dynamics  with a clear distinction between  heat and work.
Sekimoto and Sasa have demonstrated the minimum work principle 
and defined the free energy from this principle. \cite{KS} 
Their argument also includes a complementary relation which 
defines a new thermodynamic function of two state variables.  
Recently, Sekimoto and Oono have constructed
an example of steady state thermodynamics by analyzing a Langevin
dynamical model.\cite{SO}

The minimum work principle can be formulated through an equality
proposed by Jarzinski.\cite{Jarz} This equality may provide a method
to find inequalities  related to thermodynamics. 
In fact, Hatano has proved  a Jarzinski-type equality for the
transition between steady states  under a certain condition 
and has derived an inequality related to the steady state 
thermodynamics.\cite{Hatano}

As discussed by Crooks,\cite{Crooks} the Jarzinski's equality is also 
related to the fluctuation theorem proposed by Evans, Cohen and 
Morriss.\cite{Evans}  The fluctuation theorem claims a peculiar 
property of the probability of the finite time average of the 
entropy production in a non-equilibrium steady state. Gallovotti 
and Cohen have presented a mathematical proof of the fluctuation 
theorem based on the assumption that the steady state measure is 
given by the dynamical measure.\cite{Gall}
Since that time, it has been shown that the fluctuation theorem 
holds even in  stochastic systems.\cite{Kurchan,LS} 
On general grounds, Maes has presented an argument that
the fluctuation theorem can be understood by a Gibbs property
of the space-time measure.\cite{Maes}  

Transportation coefficients in non-equilibrium steady states 
have been expressed in terms of dynamical system quantities. There are
two different approaches for this. In one approach, the viscosity is related
to the sum of all Lyapunov exponents in a Hamiltonian system supplemented
with a deterministic thermostatting force.\cite{Posch,ECM}  
The other approach applies to Hamiltonian systems with open boundary 
conditions. Here, the diffusion constant is related to
the escape rate which is obtained in terms of the difference
between the sum of the positive Lyapunov exponents and the Kolmogorov-Sinai 
entropy.\cite{Gas}

Finally, we mention some  recent developments in the understanding 
of thermodynamics.  Lieb and Yngvason wrote an important paper
on axiomatic thermodynamics.\cite{Lieb} They have given an
explicit expression of the thermodynamic entropy based on a set  
axioms concerning the  adiabatic accessibility, and have proven the entropy 
principle, the second law of thermodynamics.  
Although their formulation is fully mathematical, the idea of
the explicit expression of thermodynamic entropy can be translated
into standard energetic thermodynamics.\cite{Sato,Sbook,Tbook} 

\subsection{Outline of the paper} 

This paper consists of nine sections, each of which consists of 
several subsections. In order to give a self-contained explanation,
we include  a review of thermodynamics, Hamiltonian systems, 
Boltzmann entropy, and Lyapunov analysis.  Some of these 
are no doubt rather well-known topics to specialists. However, there 
are not a large number of people who  understand all of them well. 
Also, it was our intention to write this paper so that it can be
understood by non-specialists, who have 
interest in the relation between  thermodynamic irreversibility 
and dynamical systems. The organization of the paper is  summarized below.

In Section 2 we start with a review of thermodynamics in an adiabatic 
environment. Thermodynamic irreversibility is precisely defined
based on  basic notions such as state and process. The essence of the
thermodynamic entropy is described by the entropy principle.\cite{Lieb} 
We then  explain the reason why  a Hamiltonian  system with a time-dependent
parameter  provides a model for a thermodynamic system in an adiabatic 
environment. We assume the microcanonical measure for the initial conditions
and that the systems possess the mixing property with 
respect to the measure. We also assume the existence of  certain 
large deviation properties in order to establish correspondence with the 
extensivity of thermodynamics.\cite{LD} Based on these assumptions, 
we define the equilibrium state and  most probable process in Hamiltonian 
systems. After these preliminaries, we address a main question.

In Section 3  we first review the Boltzmann formula in statistical
mechanics. In particular, defining the time-dependent Boltzmann entropy,
we derive a simple form of the Boltzmann entropy change 
for general processes. Using this formula, we calculate the  average of 
the Boltzmann entropy change for a step process, where the average
is taken over the initial conditions sampled from the microcanonical 
ensemble on an energy surface. We show that the average value is
positive in the thermodynamic limit.  We further  find that the average
value is related to the fluctuation of the Boltzmann entropy change.

In Section 4 we review the Lyapunov analysis, which is a standard 
method to study chaotic dynamical systems with numerical experiments. 
We start with the Gram-Schmidt decomposition, because it is the easiest 
computational technique for the Lyapunov analysis.\cite{SN}  
We then discuss the convergence property of an orthogonal frame. 
Since  the orthogonal frame obtained from convergence does not satisfy 
the transitive property, we define  Lyapunov vectors from the orthogonal 
frame so that this property is satisfied.\cite{GD}
Based on these Lyapunov vectors, we define  Lyapunov exponents, 
local expansion ratios, and the information loss rate.
In order to recover the symmetry of unstable and stable directions,
we also define contraction ratios. We then prove a relation between
the expansion and contraction ratios. We also derive an expression for
the weight on trajectory segments.

In Section 5 we discuss the reversibility of Hamiltonian systems.
We relate  the evolution map, Lyapunov vectors, and local expansion 
rates  for time-reversed systems to those for the original system.
The reversibility leads to a reversibility paradox. \cite{Ehren}
In order to resolve the paradox, we  need to consider the measure
for a set of the initial conditions for time-reversed systems. 
This consideration leads to  a reversibility relation expressed 
in terms of probability. 

In Section 6 we begin with the definition of {\it irreversible 
information loss}. Using the reversibility relation mentioned above,
we prove that  the irreversible information loss averaged over
the initial conditions is always non-negative.
We define the most probable value of the irreversible information 
loss in the thermodynamic limit, and we present an argument that 
this most probable value is equal to the Boltzmann entropy change. 

In Section 7  we define a quantity we call {\it excess 
information loss}, because  this quantity is more tractable than
the irreversible information loss.  We present a relation between 
the Boltzmann entropy change  and the excess information loss
based on the assumption that the  reversible part of the excess 
information loss is equal to the quasi-static excess information loss. 
This relation is identical to an equality proposed in a previous 
paper.\cite{SK}  
Furthermore, we briefly discuss a minimum excess information loss 
principle, which may be analogous in some sense to the minimum work 
principle in thermodynamics with  an isothermal environment. 
We also explain the origin of the quasi-static excess information 
loss  using  Lyapunov analysis. 

In Section 8  we report results of numerical experiments on 
a Fermi-Pasta-Ulam model\cite{FPU} with a time-dependent nonlinear term.
With these, we numerically check the assumptions in the arguments given 
in the previous sections, and numerically demonstrate several properties 
of certain quantities such as the Lyapunov exponents and Boltzmann entropy 
changes in this model. As the main numerical experiment, we confirm the 
relation between the Boltzmann entropy change  and the excess information 
loss. 

The final section is devoted to concluding remark.

\subsection{Remarks}

Our theoretical arguments include some non-rigorous, but intuitively 
reasonable statements. To as great an extent as possible, we state 
explicitly when the assertions we make are assumptions. There is one 
exception, however. We often use the expression $o(N)$ to represent 
a quantity of negligible magnitude compared to $N$ in the limit 
$N \rightarrow \infty$. This constitutes an order estimate valid in the
case that the system satisfies an appropriate condition. However, 
we do not discuss what this condition is, nor do we explicitly state 
that an assumption is involved when we neglect such a quantity. 
We simply expect that the condition is satisfied unless
an abnormal situation occurs.

We use the same font for  numbers and vectors.
We believe that the difference can be understood in the context.
Also,  a matrix is expressed as ${\cal A}$, and $ A_{ij}$ denotes 
the $(ij)$-element of this matrix.

\section{Preliminary}

In this section, we review thermodynamics and Hamiltonian
systems. We clarify basic assumptions of our theory and 
address the main question of this paper. 

\subsection{Thermodynamic irreversibility}

A thermodynamic system is characterized  by the internal energy $U$
and a set of work variables $\{ X_i \}$. 
When the value of $X_i$ is changed externally,  the energy change
is induced. The infinitely small response $dU$ is written as
\begin{equation}
dU= \sum_{i}  Y_i d X_i.
\label{adien}
\end{equation}
In thermodynamics, $X_i$ is chosen as an extensive or intensive
variable. Since the internal energy $U$ is an extensive variable,
$Y_i$ is an intensive or extensive variable, respectively.
The relation Eq. (\ref{adien}) is valid for the case that the system 
is enclosed by adiabatic walls. More formally, Eq. (\ref{adien}) 
should be regarded as a mathematical expression of a physical situation 
that the system is placed in  an adiabatic environment. 

The equilibrium state $\Sigma$ is assumed to be realized 
when the system is left  for a sufficiently long time after 
values of the work variables are fixed. This assumption
provides the operational definition of the equilibrium state.
Also, the equilibrium state $\Sigma$ is assumed to be determined 
uniquely by a set of the values of $(U,\{X_i\})$. That is, the state 
$\Sigma$ is identified with $(U,\{X_i\})$. 
When the value of $X_i$ is changed externally, a transition 
from an equilibrium state $\Sigma_0$ to another one $\Sigma_1$
occurs.  
This transition, which is denoted by $\Sigma_0 \tps \Sigma_1$,
is called the {\it thermodynamic process} or simply the {\it 
process}. More precisely, the process  is called the adiabatic process
realized in an adiabatic environment. However, in the argument 
below, we use the term process instead of adiabatic process.

Let $\Sigma_0$ and $\Sigma_1$ be arbitrary equilibrium states. 
We then ask  whether or not processes $\Sigma_0 \tps \Sigma_1$
and $\Sigma_1 \tps \Sigma_0$ are realizable. When both the processes 
are realizable, these are called  {\it reversible processes}. 
When only one process  $\Sigma_0 \tps  \Sigma_1$ is realizable, 
this process  is called the {\it irreversible process}. 
We can easily see that the process 
\begin{equation}
(U,\{X_i\}) \tps (U', \{X_i\})
\end{equation}
provides an example of irreversible processes when $U' > U$.

The essence of the thermodynamic entropy is  described by the
entropy principle:\cite{Lieb}

There exists an extensive variable $S$ given by a 
function of $\Sigma$ such that  the inequality 
\begin{equation}
S(\Sigma_1) \ge S(\Sigma_0)
\label{epri}
\end{equation}
is satisfied  if and only if a process $\Sigma_0 \tps \Sigma_1$
is realizable.  The extensive variable $S$ is  determined 
uniquely up to multiplicative and additive arbitrary constants. 

Lieb and Yngvason have proved the entropy principle 
based on the axioms concerning the  adiabatic accessibility.
\cite{Lieb} Also, in conventional thermodynamics based on work and heat, 
the entropy principle can be proved on some physical assumptions.
\cite{Sbook,Tbook}

\subsection{Hamiltonian systems}

A Hamiltonian system is characterized by a Hamiltonian function 
$H(\Gamma)$, where $\Gamma$ is  a set of canonical coordinates 
$\{q_i\}$  and  momentums $\{p_i\}$
\begin{eqnarray}
\Gamma_i &=& q_i, \\
\Gamma_{N+i} &=& p_i,
\end{eqnarray}
where $1 \le i \le N$. Equations of motion for $q_i$ and $p_i$ 
are given by
\begin{eqnarray}
\der{q_i}{t} &=& \pder{H}{p_i}, \\
\der{p_i}{t} &=& -\pder{H}{q_i}.
\end{eqnarray}
These equations are formally written as
\begin{equation}
\der{\Gamma}{t}=-{\cal J}\pder{H}{\Gamma},
\label{bigeq}
\end{equation}
where ${\cal J}$ is a $2N \times 2N$ anti-symmetric matrix
which satisfies
\begin{equation}
{\cal J}^2=-1.
\label{bigJ}
\end{equation}
Under an initial condition $\Gamma(0)$  given at $t=0$,
the phase space point at  time $t$, $\Gamma(t)$, 
is determined by the equation of motion. 

In this paper, we are concerned with Hamiltonian systems with 
a time-dependent parameter $\alpha$. 
The energy of the system $E$ at time $t$ is given by
\begin{equation}
E(t)=H(\Gamma(t),\alpha(t)).
\end{equation}  
We then  obtain the equality 
\begin{eqnarray}
\der{E}{t} &=& \sum_{i=1}^{N}
(\pder{H}{q_i}\der{q_i}{t}+\pder{H}{p_i}\der{p_i}{t})
+\pder{H}{\alpha}\der{\alpha}{t},\\
&=& \pder{H}{\alpha}\der{\alpha}{t},
\label{edert}
\end{eqnarray}
where we have used the equations of motion.
This equality is rewritten as
\begin{equation}
dE= A d\alpha,
\label{dE-law}
\end{equation}
where 
\begin{equation}
A=\pder{H}{\alpha}.
\end{equation}
Comparing Eq. (\ref{dE-law}) with  Eq. (\ref{adien}),
we find that $E$ and $\alpha$ correspond to the internal 
energy $U$ and  a work variable $X$. 
This suggests  that a Hamiltonian system with 
a time-depending parameter can be a  dynamical system model for 
thermodynamics in an adiabatic environment. 
We proceed to our discussion based on this 
expectation and attempt to find necessary issues so as to
establish consistency with thermodynamics.  

Since we are particularly interested in thermodynamic processes,
we assume that the value of $\alpha$ is  changed in a  
finite time interval $[\tau_i,\tau_f]$, that is, 
\begin{equation}
\der{\alpha(t)}{t}=0 
\end{equation}
when $t \not \in [\tau_i,\tau_f]$.
In the argument below, we assume the condition
\begin{equation}
0 \ll \tau_i \le \tau_f
\label{acon}
\end{equation}  
without an explicit remark. Note that $\ll$ in Eq. (\ref{acon}) has
been assumed for a technical reason. 
We also represent the protocol of the parameter change by $\alpha()$.

\subsection{Measure}

We assume that the initial condition given at $t=0$ is sampled from
the microcanonical ensemble on an energy surface $\Sigma$.
The measure for the ensemble is given by the microcanonical measure
\begin{equation}
\mu_{\rm mc}(d\Gamma;\Sigma) ={1\over |\Sigma|}
 {1 \over |\nabla_\Gamma H|}d\sigma,
\end{equation}
where $d\sigma$ is the Lebesgue measure on the energy
surface, and $|\Sigma|$ is given by
\begin{equation}
|\Sigma| = \int d\sigma  {1 \over |\nabla_\Gamma H|}.
\end{equation}
$\mu_{\rm mc}(\Gamma;\Sigma)$ and $|\Sigma|$ are rewritten as
\begin{eqnarray}
\mu_{\rm mc}(d\Gamma;\Sigma) &=& {1\over |\Sigma|}
    d\Gamma \delta (H(\Gamma)-E), 
\label{dfun}\\
|\Sigma| &=& \int d\Gamma \delta (H(\Gamma)-E),
\end{eqnarray}
where $d\Gamma$ is the $2N$-dimensional Lebesgue volume element in the
phase space. We also notice that Eq. (\ref{dfun}) is given by
\begin{equation}
\mu_{\rm mc}(\Delta_\epsilon(\Gamma);\Sigma) =
\lim_{\delta E \rightarrow 0} 
{\mu_L( \Delta_\epsilon(\Gamma) \cap \Sigma\circ \delta E) \over
 \mu_L( \Sigma\circ \delta E)}
\end{equation}
where $\Sigma \circ \delta E$ denotes a union of energy surfaces from
$E$ to $E+\delta E$, $\Delta_\epsilon(\Gamma)$ is a region with a
size $\epsilon$ which includes a point $\Gamma$, and  
$\mu_{\rm L}$ is the 
$2N$ dimensional Lebesgue measure.

We also assume that the systems in question  possess are ergodic and 
process the mixing property. Here, a system is called ergodic 
with respect to the microcanonical measure, when
the equality 
\begin{equation}
\lim_{T \rightarrow \infty} {1\over T}\int_0^T dt f(\Gamma(t))=
\int \mu_{\rm mc}(d\Gamma;\Sigma)f(\Gamma)
\end{equation}
holds for an arbitrary measurable function $f$ and almost all
initial conditions $\Gamma(0)$ with respect to the measure.
The mixing property with respect to the microcanonical measure
means  that  the equality 
\begin{equation}
\lim_{t \rightarrow \infty} 
\int \mu_{\rm mc}(d\Gamma(0);\Sigma)f(\Gamma(0))g(\Gamma(t))
= \int \mu_{\rm mc}(d\Gamma;\Sigma)f(\Gamma)
\int \mu_{\rm mc}(d\Gamma;\Sigma)g(\Gamma)
\end{equation}
holds for arbitrary measurable functions $f$ and $g$.
It is easily proved that a mixing system possesses the ergodicity.

Suppose that the initial condition at  time $t=-t'$, $t' > 0$, is 
sampled from an ensemble with the measure 
\begin{equation}
\mu_{ f}(d\Gamma(-t');\Sigma)=
\mu_{\rm mc}(d\Gamma(-t');\Sigma)f(\Gamma(-t')),
\end{equation}
where $f$ is a measurable function normalized in such a
way that 
\begin{equation}
\int_\Sigma \mu_{\rm mc}(d\Gamma(-t');\Sigma)f(\Gamma(-t'))=1.
\end{equation}
Then, the mixing property leads to
\begin{equation}
\lim_{t' \rightarrow \infty} 
\int \mu_{ f}(d\Gamma(-t');\Sigma) g(\Gamma(0)) 
= 
\int \mu_{\rm mc}(d\Gamma ;\Sigma)g(\Gamma).
\label{mixf}
\end{equation}
That is,  the average of $g(\Gamma(0))$ with respect to
$\mu_{f}(d\Gamma(-t');\Sigma)$ is the same as the average 
of $g(\Gamma(0))$ with respect to  $ \mu_{\rm mc}(d\Gamma(0);\Sigma)$ 
when $t' \rightarrow \infty$.
Using Eq. (\ref{mixf}), we check numerically the validity of the mixing 
property and we can prepare the microcanonical ensemble at $t=0$
in the following way.  

First, we prepare a set of  the initial conditions at $t=-t'$  sampled  
from an ensemble with a measure  absolutely continuous with respect 
to the Lebesgue measure on the energy surface $\Sigma$.  (It can be
done easily in numerical experiments.) Then, we take an average of 
a dynamical variable, for example $A$, at  $t=0$.  We carry out two 
experiments for two different measures assumed at $t=-t'$. If the 
average values  coincide, we may regard that the system possesses the 
mixing property.
\footnote{However, precisely speaking, this is nothing but a 
confirmation of one of necessary conditions for the mixing 
property.} 
Also, from Eq. (\ref{mixf}), we find that  this average value is the average 
for the microcanonical measure at $t=0$,  $\mu_{\rm mc}(d\Gamma ;\Sigma)$.
This implies that we can prepare the microcanonical ensemble at $t=0$.

\subsection{Thermodynamic limit}

In thermodynamics,  the internal energy $U$ is an extensive variable, 
and  a work variable is an extensive or intensive variable. 
In order to establish the consistency with thermodynamics, we assume the 
following  large deviation property\cite{LD} which may be 
closely related 
to the extensivity of the energy:

Let $\Pi_E(E_1) dE$ be a probability that final energy 
after the parameter change takes a value in the region $[E_1, E_1+dE]$.
Then, $\Pi_E$ is written in the form
\begin{equation}
\Pi_E(E_1) \sim \exp(-N \phi_E(E_1/N)),
\label{LDene}
\end{equation}
in the appropriate asymptotic limit including $N \rightarrow \infty$.

Several remarks are mentioned. (i) The appropriate limit in 
Eq. (\ref{LDene}) is called  the {\it thermodynamic limit}.
In the argument below, the limit $N \rightarrow \infty$ always implies
the thermodynamic limit without an explicit remark.
(ii) The probability of final energy is induced from the measure for 
the ensemble of the initial conditions. (iii) $\phi_E$ is called
a rate function, and is a non-negative convex function with zero.
The zero of $\phi_E$, $ \bar E_{1*}$,  is called the most probable 
value of $E_1/N$. 

We next discuss the extensivity or intensivity  of  $\alpha$.
We pay attention to the case that $\alpha$ is an intensive parameter.
(The reversed case is similarly discussed.)  The variable $A$ then
turns out to be an extensive variable, which is characterized by 
the large deviation property:

Let $\Pi_A(A') dA$  be a probability that $A$ takes a value
in the region $[A', A'+dA]$ at $t=0$. Then, $\Pi_A$ is written
in the form 
\begin{equation}
\Pi_A(A') \sim \exp(-N \phi_A(A'/N)),
\label{LDA}
\end{equation}
in the thermodynamic limit.

Note that the probability density $\Pi_A$ is determined 
by the measure for the ensemble.  Since the most probable 
value  of $A/N$, $\bar A_*$, exists for each energy surface, 
we write $\bar A_*(\Sigma)$ when we emphasize the state dependence.

\subsection{Equilibrium state}

We assume that  the equilibrium state in thermodynamics 
corresponds uniquely to the energy surface. That is the reason 
why we used  the same symbol $\Sigma$ of an energy surface in 
Section 2.3   as the equilibrium state in Section 2.1.  Also, the energy 
surface is specified 
by a set of quantities $(E,\alpha)$.  In the argument below, $\Sigma$
denotes an equilibrium state, an energy surface and a set of
quantities $(E,\alpha)$.

Let us discuss a condition under which we can know whether or not 
the equilibrium state is realized. The term 'equilibrium' implies 
that nothing changes due to the balance.
Thus, it is natural to find a quantity which  does not change  at 
equilibrium. Although the energy $E$ does not change when 
$t \ge \tau_f$, it is strange  that the equilibrium state is realized 
immediately after the parameter change is finished. 
The energy cannot be used as an indicator of the equilibrium
state. The next candidate of the indicator may be the variable $A$.
However, since the trajectory never converges to a fixed point,
the value of $A$  remains  time-dependent. We then notice here that
the argument for the nature of equilibrium  should be developed 
with the thermodynamic limit.

Suppose that $\Gamma(\tau_f) \in \Sigma_1$. In general,
$A(\tau_f)/N$ is not equal to $\bar A_*(\Sigma_1)$.
However, from large deviation and mixing properties, we can expect
\begin{equation}
\lim_{t \rightarrow \infty} 
{A(t) \over N} \rightarrow  \bar A_*(\Sigma_1)
\label{n-relax}
\end{equation}
in the thermodynamic limit.
When $A(t)/N$ is sufficiently close to 
$ \bar A_*(\Sigma_1)$ up to a certain time,
\footnote{Formally, we should consider the limit $N \rightarrow \infty$
before the limit  $t \rightarrow \infty$. In the experiment with  finite $N$,
we should focus on an asymptotic regime up to a certain time.}
we assume that the state is at the equilibrium.
There may be other important physical quantities to be checked. 
However, since we do not have any criteria for the importance, 
we assume that the relaxation of the variable $A$ is enough to 
identify the  equilibrium state.

\subsection{Most probable process}

Suppose that an equilibrium state $\Sigma_0$ is realized 
at $t=0$ and that another equilibrium state is realized in an 
energy surface $\Sigma_1$ after a sufficiently long time from 
$t=\tau_f$.  We call this transition  a process in the
similar way as thermodynamics. However, since $\Sigma_1$ depends on 
$\Gamma(0)$, $\Sigma_1$ is not determined uniquely  when we assume 
the initial energy surface $\Sigma_0$ and the protocol of the 
parameter change $\alpha()$. Here, in order to establish the 
correspondence with thermodynamics, we assume that the large 
deviation property of the path of $E$:

Let $\Pi_{E:{\rm path}}(\{ E'(t), 0 \le t' \le \tau \}) \prod_t dE(t) $
be a path probability  that $E(t)$ takes a value in the region 
$[E', E'+dE(t)]$ at  time $t$. Then, $\Pi_{E:{\rm path}}$ 
is written in the form 
\begin{equation}
\Pi_{E {\rm :path}}(\{ E'(t), 0 \le t' \le \tau \} ) 
\sim \exp(-N \phi_{E{\rm :path}}(\{ E'(t)/N, 0 \le t' \le \tau \} ),
\label{LDpath}
\end{equation}
in the thermodynamic limit. 

The probability density $\Pi_{E {\rm :path}}$ is determined 
by the measure for  the ensemble of the initial conditions. The rate function 
$\phi_{E {\rm :path}}$ is a function of path segments 
$\{ E'(t)/N, 0 \le t' \le \tau \}$, and there is a most probable 
path $\{ \bar E_{*}(t'), 0 \le t' \le \tau \}$ which minimizes 
the rate function. 
Then, since the  parameter is changed in a deterministic way, 
the most probable process is defined as
$ \{ (N\bar E_*(t'),\alpha(t')) ,\ 0 \le t' \le \tau_f\}$. 
The most probable process is denoted by 
\begin{equation}
\Sigma_0 \fwdp \Sigma_1,
\end{equation}
where $\Sigma_1=(N \bar E_{*}(\tau_f), \alpha(\tau_f))$,
and it is identified with the 
thermodynamic  process $\Sigma_0 \tps \Sigma_1$.

\subsection{Main question}

Let us summarize our basic assumptions and address the main question.
When one attempts to study thermodynamic irreversibility
in Hamiltonian systems, there seem three problems.
The first problem is related to the measure for the initial 
conditions, where, as one example, a condition which determines the most 
natural measure is concerning. The second  problem  is related 
to the reason why macroscopic variables behave in a deterministic way. 
The discussion of large deviation properties is one way to consider the
problem.

In this paper, we do not enter these problems deeply.  
As mentioned above, we assume that  Hamiltonian systems in question 
possess the mixing property with respect to the microcanonical measure,
and we also assume  the large deviation properties of $A$ and $E$ in the 
thermodynamic limit. Nevertheless, we believe that there is still 
an important problem to be solved. We ask how the thermodynamic law 
is  established.  In other word, we ask whether or not we can find a 
state variable  satisfying  the entropy principle from dynamical systems. 
Let us write the question explicitly.

Let $\Sigma_0 \fwdp \Sigma_1$ be an arbitrary most probable process.
Then, find a state variable $S$ such that 
\begin{equation}
S(\Sigma_1) \ge S(\Sigma_0),
\end{equation}
where the equality holds only when the reversed process 
$\Sigma_1 \fwdp \Sigma_0$ is realizable.

\section{Statistical mechanics}

In statistical mechanics, the thermodynamic entropy is calculated 
as the Boltzmann entropy. We then review fundamental properties 
of the Boltzmann entropy and discuss whether or not  we can 
answer to the main question by using the Boltzmann entropy. 

The thermodynamic entropy  takes a constant value along an 
arbitrary quasi-static process $\Sigma_0 \qsp \Sigma_1$, which 
is realized by infinitely slow change of the parameter value.
Then, in developing statistical mechanics, we first attempt
to find such a quantity.\cite{Toda}  The adiabatic theorem
ensures the existence of an invariant quantity along quasi-static 
processes.  We thus  start with the adiabatic theorem.

\subsection{Adiabatic theorem} 

Let $\Omega(\Sigma)$ be the phase space volume enclosed 
by an energy surface $\Sigma=(E,\alpha)$.
\begin{equation}
\Omega(E,\alpha)=\int d\Gamma \theta(E-H(\Gamma,\alpha)).
\label{omega}
\end{equation}
When the value of $\alpha$ is changed in time, the energy 
of the system changes. We define the time evolution of 
the phase space volume as
\begin{equation}
\Omega(t)=\Omega(E(t),\alpha(t)),
\end{equation}
where note that $E(t)$ depends on $\Gamma(0)$.
We then obtain
\begin{equation}
\der{\Omega}{t}= \left [
\pder{\Omega}{\alpha}+\pder{\Omega}{E} A \right]
\der{\alpha}{t},
\label{der-o}
\end{equation}
where we have used the equality 
\begin{equation}
\der{E}{t}=A \der{\alpha}{t},
\end{equation}
which is given by Eq. (\ref{edert}). 

By using the expression Eq.(\ref{omega}), we derive
\begin{eqnarray}
\pder{\Omega}{E} &=& |\Sigma|,
\label{derE} \\
\pder{\Omega}{\alpha} &=& -|\Sigma|
\bra A \ket_{\Sigma},
\label{dera}
\end{eqnarray}
where $|\Sigma|$ and $ \bra f  \ket_{\Sigma} $
are defined as 
\begin{eqnarray}
|\Sigma| &=& \int d\Gamma \delta (E-H(\Gamma,\alpha)), \\
 \bra f  \ket_{\Sigma} &=&
{1 \over |\Sigma|}
\int_{\Sigma}  { d \sigma \over |\nabla_{\Gamma} H|}f(\Gamma).
\end{eqnarray}
$\bra f \ket_{\Sigma}$ corresponds to the average of $f$ over the 
micro-canonical ensemble on the energy surface  $\Sigma$. 
Substituting Eqs. (\ref{derE}) and (\ref{dera}) into
Eq. (\ref{der-o}), we obtain
\begin{eqnarray}
\der{\Omega}{t} &=& |\Sigma| \left [A -\bra A\ket_{\Sigma} \right ]
\der{\alpha}{t},
\label{der-o2} \\
&=&  |\Sigma| \delta A
\der{\alpha}{t},
\label{der-o3}
\end{eqnarray}
where we have defined a new variable $\delta A$ as
\begin{equation}
\delta A= A -\bra A \ket_{\Sigma}.
\end{equation}


We now prove the adiabatic theorem which states that  the equality
\begin{equation}
\Omega(\Sigma_0)=\Omega(\Sigma_1)
\label{adigoal}
\end{equation}
holds for an arbitrary quasi-static process
$ \Sigma_0 \qsp \Sigma_1$.

(proof)

Put $\Sigma_0=(E_0,\alpha_0)$ and $\Sigma_1=(E_\infty,\alpha_\infty)$.
We decompose the quasi-static process into $n$ quasi-static 
processes such that 
\begin{equation}
(\alpha_j,E_j) \qsp  (\alpha_{j+1}, E_{j+1}),
\end{equation}
where $0 \le j \le n-1$, and $\alpha_{j+1}=\alpha_j+\Delta \alpha$
with 
\begin{equation}
\Delta \alpha={\alpha_\infty-\alpha_0 \over n}.
\end{equation}
Note that $\alpha_n=\alpha_{\infty}$ and $E_n=E_{\infty}$.
We first assume 
\begin{eqnarray}
\Delta \Omega_j &=&
\Omega(E_{j+1},\alpha_{j+1})-\Omega(E_{j},\alpha_{j}),\\
&=& O((\Delta \alpha)^2)
\label{goal1}
\end{eqnarray}
for large  $n$. We then obtain 
\begin{eqnarray}
\Omega(E_0,\alpha_0)-\Omega(E_\infty,\alpha_\infty)
&=& 
\lim_{n\rightarrow \infty} \sum_{j=1}^{n-1} O( (\Delta \alpha)^2), \\
&=& \lim_{n\rightarrow \infty}O({1\over n}),\\
&=& 0.
\end{eqnarray}
This shows the adiabatic theorem. 

We next prove Eq. (\ref{goal1}). 
Without loss of generality, we can assume that the value of $\alpha$
is monotonically changed  from $\alpha_j$ to $\alpha_j+\Delta \alpha$. 
Defining the protocol of the parameter change $\alpha_\tau()$ as 
\begin{equation}
\alpha_\tau(t)= \alpha_j+(\Delta \alpha){t \over \tau},
\end{equation}
we calculate $\Delta \Omega_j$ from Eq. (\ref{der-o3}) in the 
following way:
\begin{eqnarray}
\Delta \Omega_j &=& \lim_{\tau \rightarrow \infty}
\int_{0}^\tau dt
|\Sigma(t)|\delta A(t)\der{\alpha_\tau}{t}, \\
&=& \lim_{\tau \rightarrow \infty}
{(\Delta \alpha) \over \tau} \int_{0}^\tau dt
|\Sigma(t)|\delta A(t).
\label{qss}
\end{eqnarray}
When $\Delta \alpha$ is sufficiently small, $\Delta \Omega_j$
is evaluated as 
\begin{eqnarray}
\Delta \Omega_j &=& |\Sigma(0)| \lim_{\tau \rightarrow \infty}
{(\Delta \alpha) \over \tau} \int_{0}^\tau dt
\delta A(t) +O((\Delta \alpha)^2), \\
&=& O((\Delta \alpha)^2),
\label{qss2}
\end{eqnarray}
where  we have used 
\begin{equation}
\lim_{\tau \rightarrow \infty}
{1\over \tau} \int_{0}^\tau dt
\delta A(t) =0,
\end{equation}
which is  equivalent to
\begin{equation}
\lim_{\tau \rightarrow \infty} {1\over \tau} \int_{0}^\tau dt
A(t) = \bra A \ket_{\Sigma}.
\end{equation}
This equality holds for almost all initial conditions
with respect to the Lebesgue measure on the energy surface
because of the ergodicity with respect to the microcanonical measure.

\hfill (q.e.d)

\subsection{Boltzmann entropy}

We define the  Boltzmann entropy $S_{\rm B}$ as
\footnote{See page 371 in the Boltzmann's book\cite{Boltz} as an
explicit presentation of the Boltzmann formula. However, 
the expression of Eq. (\ref{bolt}) was first proposed by Gibbs as 
the correspondence of thermodynamic entropy. See page 128 in the Gibbs's book.
\cite{Gibbs}  The monograph by P. and T. Ehrenfest is also useful to
know  contemporary ideas with them.\cite{Ehren}}
\begin{equation}
S_{\rm B}(\Sigma)=\log \Omega(\Sigma),
\label{bolt}
\end{equation}
where the Boltzmann constant is assumed to be the unity.
For later convenience, we  define  the temperature
$T(\Sigma)$ as
\begin{equation}
T(\Sigma) = \left( \pder{S_{\rm B}}{E} \right)^{-1}
={\Omega(\Sigma) \over |\Sigma|} .
\label{ondo}
\end{equation}
Although Eq. (\ref{bolt}) is the formula which makes us possible to 
calculate the thermodynamic entropy for the equilibrium state $\Sigma$, 
we  define the time evolution of $S_{\rm B}$ as
\footnote{Do not confuse it with the time evolution of the
$H$-function in the $H$-theorem of Boltzmann.\cite{Ehren,Boltz}}
\begin{equation}
S_{\rm B}(t)=S_{\rm B}(\Sigma(t)).
\end{equation}
Then, from Eqs. (\ref{der-o3}), (\ref{bolt}) and (\ref{ondo}),
we obtain
\begin{equation}
\der{S_{\rm B}}{t} =  {1 \over T(\Sigma)} \delta A\der{\alpha}{t}.
\end{equation}
The integration of this equation during the time interval 
$\left [0,\tau\right ]$ leads to 
\begin{equation}
\Delta S_{\rm B}= \int_{0}^\tau dt
{\delta A(t) \over T(\Sigma(t))}\der{\alpha}{t}.
\label{final}
\end{equation}

Here, we  notice that $A d\alpha$  equals  to the energy change
$\Delta E$ during a time interval $\left [t,t+dt\right ]$ and that 
$ \bra A \ket_{\Sigma}d\alpha$ may be interpreted as the 
quasi-static work, $W_{\rm qs}$,  calculated under 
the condition that the system stays {\it virtually} 
in the energy surface.  $W_{\rm qs}$ is identical to
the work done in the actual process which can be realized
when the system contacts a heat bath with
slowly changing temperature. The quasi-static heat $Q_{\rm qs}$ from the
heat bath is then  given by
\begin{equation}
Q_{\rm qs}=\Delta E-W_{\rm qs}=\delta A d\alpha.
\end{equation}
Using $Q_{\rm qs}$,  we rewrite Eq. (\ref{final}) as
\begin{eqnarray}
\Delta S_{\rm B} &=& \int_{0}^\tau dt {1 \over T}\der{Q_{\rm qs}}{t},\\
         &=& \int_{0}^\tau {dQ_{\rm qs} \over T}.
\label{final:2}
\end{eqnarray}
This should be compared with the Clausius's formula
\begin{equation}
\Delta S= \int {d'Q \over T},
\label{carnot}
\end{equation}
where $d'Q$ is an infinitely small quasi-static heat 
exported from a heat bath. In this way, the Boltzmann entropy
turns out to be identified to the thermodynamic entropy.

\subsection{Entropy change for step processes}

We discuss the step process given  by
\begin{equation}
\der{\alpha}{t}=\delta(t) \Delta \alpha.
\label{stepp}
\end{equation}
In the argument below, $\Sigma_0=(E_0,\alpha_0)$ and  
$\Sigma_1=(E_1,\alpha_1)$ denote the initial and final states, 
respectively.
By substituting Eq. (\ref{stepp}) into Eq. (\ref{final}),
we have
\begin{equation}
\Delta S_{\rm B} =  {1 \over 2} 
\left[{\delta A (0_+)  \over T(\Sigma_1)}
+ {\delta A (0_-)  \over T(\Sigma_0)}\right]
\Delta \alpha,
\label{echange}
\end{equation}
where note that $\delta A$ and $T$ are discontinuous at $t=0$.
We consider  the average over initial conditions sampled from
the microcanonical ensemble on the energy surface $\Sigma_0$. 
This average is denoted by $\bra \ \ket_0$. We  calculate
$\bra S_{\rm B} \ket_0 $ as 
\begin{equation}
 \bra \Delta S_{\rm B}  \ket_0 =
{(\Delta \alpha )^2 \over 2T_0}
\left[{1 \over \Sigma} \pder{\Sigma}{E} \bra (\delta A)^2 \ket_0 
+ {\partial \over \partial E} \bra (\delta A)^2 \ket_0 \right]
+o((\Delta \alpha)^2),
\label{ent:step}
\end{equation}
where $T_0=T(\Sigma_0)$. (The proof will be given below.)
Further, since the evaluation 
\begin{equation}
{\partial \over \partial E} 
\bra (\delta A)^2 \ket_0  = o(N)
\end{equation}
is expected when $N \rightarrow \infty$, 
we obtain
\begin{equation}
 \bra \Delta S_{\rm B}  \ket_0 =
{(\Delta \alpha )^2 \over 2T_0^2}
 \bra (\delta A)^2 \ket_0 +o(N,(\Delta \alpha)^2) >0,
\label{ent:step:N}
\end{equation}
where we have  used 
\begin{equation}
{1 \over \Sigma} \pder{\Sigma}{E} = {1 \over T}+O({1\over N} ).
\end{equation}

In the thermodynamic limit, $\bra \Delta S_{\rm B} \ket_0$ is equal
to the entropy difference $S_{\rm B}(\Sigma_1)-S_{\rm B}(\Sigma_0)$
for the most probable process $\Sigma_0 \fwdp \Sigma_1$. Thus,
we conclude 
\begin{equation}
S_{\rm B}(\Sigma_1) \ge S_{\rm B}(\Sigma_0)+o(N,(\Delta \alpha)^2)
\end{equation}
for the most probable step process $\Sigma_0 \fwdp \Sigma_1$.

Furthermore, from Eq. (\ref{echange}), the fluctuation 
$\bra (\Delta S_{\rm B})^2 \ket_0$ is calculated as 
\begin{equation}
\bra (\Delta S_{\rm B})^2 \ket_0= {\bra (\delta A)^2 \ket_0 \over T_0^2}
(\Delta \alpha)^2+o((\Delta \alpha)^2).
\end{equation}
Combing this result with  Eq. (\ref{ent:step:N}), we obtain
the equality 
\begin{equation}
\bra \Delta S_{\rm B} \ket_0=
 {1 \over 2} \bra (\Delta S_{\rm B})^2 \ket_0
+o((\Delta \alpha)^2,N).
\label{FDT}
\end{equation}
This is the fluctuation-response relation for the entropy change. 

Now, we prove Eq. (\ref{ent:step}).

(proof)

We first  find 
\begin{equation}
\bra \left[{\delta A (0_+) \over T (\Sigma_1)} +
      {\delta A (0_-) \over T (\Sigma_0)} \right]
\ket_0
={\bra \delta A (0_+) \ket_0 \over T_0}+O((\Delta \alpha)^2).
\end{equation}
{}From Eq. (\ref{echange}), we have
\begin{equation}
 \bra \Delta S_{\rm B}  \ket_0 =
{ \bra \delta A (0_+) \ket_0  \over 2T_0}(\Delta \alpha )
+O( (\Delta \alpha)^3).
\label{237}
\end{equation}
Let us evaluate  $\bra {\delta A (0_+) } \ket_0$ up to
the order of $\Delta \alpha$, where 
\begin{equation}
\bra \delta A(0_+) \ket_0= 
\bra \pder{H}{\alpha}(\Gamma(0_+),\alpha(0_+)) \ket_0
-\bra \bra \pder{H}{\alpha} \ket_{\Sigma_1} \ket_0.
\label{defdela}
\end{equation}
Since $\Gamma(0_+)=\Gamma(0_-)$, we expand the first term
of the right-hand side in such a way that 
\begin{equation}
\pder{H}{\alpha}(\Gamma(0_+),\alpha(0_+))=
\pder{H}{\alpha}(\Gamma(0_-),\alpha(0_-))
+\pdert{H}{\alpha}(\Gamma(0_-),\alpha(0_-))\Delta \alpha
+O( (\Delta \alpha)^2).
\end{equation}
Taking the average over the initial conditions, we obtain
\begin{equation}
\bra \pder{H}{\alpha}(\Gamma(0_+) ,\alpha(0_+)) \ket_0 =
\bra \pder{H}{\alpha} \ket_0 
+\bra \pdert{H}{\alpha} \ket_0 \Delta \alpha
+O( (\Delta \alpha)^2).
\label{240}
\end{equation}

We next evaluate the second term of right-hand side of
Eq. (\ref{defdela}).
\begin{equation}
 \bra \bra \pder{H}{\alpha} \ket_{\Sigma_1} \ket_0
= \bra {1 \over |\Sigma_1|} 
 \int d\Gamma \pder{H}{\alpha} \delta(H(\Gamma,\alpha_1)-E_1)
 \ket_0. 
\end{equation}
We notice that  there are four terms which include 
$\Delta \alpha$.  (i) $\Delta \alpha$ appears in 
${1 / |\Sigma_1|} $, (ii) it  appears in  $\partial H/\partial 
{\alpha}$ in the integrand, (iii) it appears in 
$H(\Gamma,\alpha_1)$ in the Dirac's delta function,
and (iv) it appears in $E_1$ in 
the Dirac's delta function. 
We extract the contribution proportional to $\Delta \alpha$
in each term.

(i) The first term: The contribution is 
\begin{equation}
-{1 \over |\Sigma|^2} \left. \der{|\Sigma|}{\alpha} \right\vert_0
  \int d\Gamma \pder{H}{\alpha} \delta(H(\Gamma,\alpha_0)-E_0),
\label{1con0}
\end{equation}
where we have defined 
\begin{equation}
{d \over d\alpha}={\partial \over \partial \alpha}
+A {\partial \over \partial E}.
\end{equation}
We here note that the equality 
\begin{equation}
\der{\Omega}{\alpha}=0.
\label{ident}
\end{equation}
holds owing to the adiabatic theorem. This equality and 
the relation $\Omega=|\Sigma| T$ leads to
\begin{equation}
\left. \der{|\Sigma|}{\alpha} \right\vert_0=-{|\Sigma_0| \over T_0} 
\left. \der{T}{\alpha}\right\vert_0 .
\end{equation}
Thus,  Eq. (\ref{1con0}) becomes
\begin{equation}
{ 1 \over T_0} \left.\pder{T}{\alpha}\right\vert_0 \bra A \ket_0 .
\label{1con}
\end{equation}

(2) The second term: Without any calculation, 
the contribution from the second term is 
\begin{equation}
\bra \pdert{H}{\alpha} \ket_0 .
\label{2con}
\end{equation}

(3) The third and forth terms: In deriving the third and forth 
terms, we employ the following formula
\begin{eqnarray}
\int d\Gamma \delta'(H-E) f(\Gamma)
&=& \int dE' \int_{H=E'}  {d\sigma \over |\nabla_\Gamma H|}
\delta'(E'-E) f(\Gamma), \\
&=& 
-{\partial \over \partial E'} \left. \left[
\int_{H=E'}  {d\sigma \over |\nabla_\Gamma H| } f(\Gamma)\right]
\right\vert_{E'=E}, \\
&=& 
-{\partial \over \partial E'} \left.\left[
\bra f \ket_{(E',\alpha)} |\Sigma(E',\alpha)| 
\right]\right\vert_{E'=E} .
\end{eqnarray}
Owing to this formula, we calculate the contribution 
from the third term
\begin{eqnarray}
&\phantom{=}&
{1 \over |\Sigma|_0} 
\int d\Gamma \left(\pder{H}{\alpha}\right)^2
\delta'(H(\Gamma,\alpha_0)-E_0)  \\
&=& -{1 \over |\Sigma|_0} {\partial \over \partial E} 
\left.\left[
 \bra \left(  \pder{H}{\alpha}\right)^2  \ket_\Sigma |\Sigma|
\right]\right\vert_0, \\
&=& - {\partial \over \partial E} 
 \left.\bra \left(  \pder{H}{\alpha}\right)^2  \ket_\Sigma\right\vert_0
-{1 \over |\Sigma|_0} \left.\pder{|\Sigma|}{E}\right\vert_0
\bra \left(  \pder{H}{\alpha}\right)^2  \ket_0 .
\label{3con}
\end{eqnarray}
Similarly, the contribution from the forth term is obtained as
\begin{eqnarray}
&\phantom{=}& -{1 \over |\Sigma|_0} 
\int d\Gamma \pder{H}{\alpha} \delta'(H(\Gamma,\alpha_0)-E_0)
\bra \pder{H}{\alpha} \ket_0 \\
&=&
{1 \over |\Sigma|_0} {\partial \over \partial E} 
\left.\left[
\bra \pder{H}{\alpha} \ket |\Sigma|
\right]\right\vert_0
\bra \pder{H}{\alpha} \ket_0, \\
&=& 
\bra \pder{H}{\alpha} \ket_0 
{\partial \over \partial E} 
\left.\left[\bra \pder{H}{\alpha} \ket \right]\right\vert_0  
+
{1 \over |\Sigma|_0} \left.\pder{|\Sigma|}{E}\right\vert_0 
 \bra \pder{H}{\alpha} \ket^2.
\label{4con}
\end{eqnarray}
The contributions from the third and forth terms are combined in 
the form
\begin{equation}
-{1 \over \Sigma_0} \left.\pder{\Sigma}{E}\right\vert_0 
\bra (\delta A)^2 \ket_0
- \left.{\partial \over \partial E} \bra (\delta A)^2 \ket_0
  \right\vert_0 
- \bra A \ket_0  \left.\pder{\bra A \ket_\Sigma }{E}\right\vert_0.
\label{34con}
\end{equation}
Then, all the contributions given by Eqs. (\ref{1con}), 
(\ref{2con}) and (\ref{34con}) are summarized as
\begin{eqnarray}
 \bra \bra \pder{H}{\alpha} \ket_{\Sigma_1} \ket_0
&=& 
\bra \pder{H}{\alpha} \ket_0 
+\bra \pdert{H}{\alpha} \ket_0 \Delta \alpha \\
&-&
{1 \over \Sigma} \pder{\Sigma}{E} 
\bra (\delta A)^2 \ket_0 \Delta \alpha - 
{\partial \over \partial E} 
\bra (\delta A)^2 \ket_0 \Delta \alpha ,
\label{res:241}
\end{eqnarray}
where we have used the equality
\begin{equation}
{ 1 \over T} \pder{T}{\alpha} -  \pder{\bra A \ket_\Sigma }{E}=0.
\label{maxwell}
\end{equation}
(The proof of this equality will be given below.)
The substitution of  Eqs. (\ref{240}) and (\ref{res:241}) to
Eq. (\ref{defdela}) yields
\begin{equation}
\bra \delta A (0_+) \ket_0={1 \over \Sigma_0} 
\left.\pder{\Sigma}{E}\right\vert_0 
\bra (\delta A)^2 \ket_0 \Delta \alpha  + 
\left.{\partial \over \partial E} 
\bra (\delta A)^2 \ket_\Sigma\right\vert_0 \Delta \alpha
+O((\Delta \alpha)^2) .
\end{equation}
Recalling Eq. (\ref{237}), we finally obtain
Eq. (\ref{ent:step}). 

\hfill (q.e.d)

Here, the proof of Eq. (\ref{maxwell}) is shown.
One may find that Eq. (\ref{maxwell}) is equivalent to a Maxwell's 
relation.

(proof)

For simplicity, we use the abbreviation $A$
for $\bra A \ket_\Sigma$. Then, $A$ is  a function of 
$(S_{\rm B},\alpha)$. Since we can write  
\begin{equation}
A=A(S_{\rm B},\alpha)=\pderf{E}{\alpha}{S_{\rm B}},
\end{equation}
we calculate
\begin{eqnarray}
\pderf{A}{E}{\alpha}&=&\pderf{A}{S_{\rm B}}{\alpha}
\pderf{S_{\rm B}}{E}{\alpha}, \\
&=& {1\over T}{\partial^2 E \over \partial S_{\rm B} \partial \alpha}, \\
&=& {1\over T}\pder{T}{\alpha}.
\end{eqnarray}

\hfill (q.e.d)

\subsection{Remark}

In this sections, we have shown  that $\Delta S_{\rm B}$ is positive 
for  most probable step processes. One may then ask whether or 
not $\Delta S_{\rm B}$ is positive  for arbitrary processes. 
As one example,  one may evaluate $\Delta S_{\rm B}$ near 
quasi-static processes based on several physical assumptions. 
However, if we consider this question from the definition 
of $\Delta S_{\rm B}$, it seems hard to  obtain any general
results. 

Nevertheless,  since $S_{\rm B}$ is equivalent to the thermodynamic entropy,
we expect the inequality 
\begin{equation}
\Delta S_{\rm B} \ge o(N)
\label{bg}
\end{equation}
holds for  most probable processes in general. 
We will discuss  the validity of Eq. (\ref{bg}) in Section 6.2
based on Lyapunov analysis of chaotic systems. We now leave
statistical mechanics and  enter  Lyapunov analysis.

\section{Lyapunov analysis}

One of essential features of chaotic systems is the sensitivity 
of initial conditions.  Consider a trajectory segment, $\{\Gamma(t),
\ 0 \le t \le \infty\}$.  Almost all trajectories starting from  
phase space points in a neighborhood  at $\Gamma(0)$ separate 
exponentially in time from the trajectory $\Gamma()$.  
Such a behavior can be discussed quantitatively by measuring 
the expansion of vectors in the tangent spaces around the trajectory.  
More generally, we can discuss the time evolution of the
$k$-dimensional 
volume element, which is given  by the exterior product of 
$k$ independent vectors in the tangent space. (See Appendix for
basic properties of the volume element and  exterior product.)
Such an argument 
includes the Liouville's theorem as a special case ($k=2N$), 
which states that the $2N$ dimensional volume element  keeps 
its volume  along the trajectory.  {}From the observation for
both the cases that  $k=1$ and $k=2N$, 
we expect that the tangent space at each point is decomposed into 
subspaces associated with the expansion  ratios. 
 
Indeed, it has been known that the multiplicative ergodic
theorem of Oseledets  provides a  mathematical description of 
the naive expectation.\cite{Oseledets} Nowadays,  the analysis of 
tangent spaces, which is often referred to as {\it Lyapunov analysis},
becomes a standard technique to discuss chaos owing to the establishment
of a numerical calculation method.\cite{SN}

In this section, we pay attention to  Hamiltonian systems without 
parameter  change except for the final two subsections, and we review 
the Lyapunov analysis with emphasising its computational aspects.  

When the value of $\alpha$ is not changed in time, 
the evolution map from $t=t_0$ to $t=t_1$ takes a form
$U_{t_1-t_0}$ and  satisfies
\begin{equation}
\Gamma(t)=U_t(\Gamma(0)).
\end{equation}
The change of the trajectory at time $t$, $\delta \Gamma(t)$, 
against infinitely small change of the initial condition, 
$\delta \Gamma(0)$, is written as
\begin{eqnarray}
\delta \Gamma(t)&=& U_t(\Gamma(0)+\delta \Gamma(0))
                 -U_t(\Gamma(0)), \\
                &=& {\cal T}(t,\Gamma(0)) \delta \Gamma(0).
\end{eqnarray}
${\cal T}(t,\Gamma(0))$ is called the linearized evolution map
and is calculated by  numerical integration of the linearized
evolution equation. Note that the matrix ${\cal T}(t,\Gamma(0))$ 
is determined by the trajectory segment 
$\{ \Gamma(t'), 0 \le t' \le t\}$. We analyze 
the matrix ${\cal T}(t,\Gamma(0))$ below.

\subsection{Gram-Schmidt decomposition}

Let $\{  {e}_i, \ 1 \le i \le 2N\} $ be 
a set of orthogonal unit vectors  given 
randomly  in the tangent space at $\Gamma(0)$.
For a while, we will use the abbreviation ${\cal T}$ for ${\cal T}
(t, \Gamma(0))$. Since almost all vectors expand toward the most unstable
direction,   the direction of the vector  
${\cal T} {e}_1$  may approach  to the most unstable 
direction when $t$ is sufficiently large.
We thus define a unit vector in the tangent space 
at $\Gamma(t)$ as
\begin{equation}
 {f}_1 = {{\cal T}  {e}_1 \over |{\cal T}  {e}_1|
}.
\end{equation}
$ {f}_1$ is expected to indicate the most unstable 
direction at $\Gamma(t)$ when $t \rightarrow \infty$. 
Similarly, we define the most unstable direction 
in the orthogonal co-space of $ {f}_1$ 
\begin{equation}
 {f}_2 = 
{ {\cal T}  {e}_2-({\cal T}  {e}_2,  {f}_1)
\over 
| {\cal T}  {e}_2-({\cal T}  {e}_2,  {f}_1) |
}.
\end{equation}
Repeating the similar consideration, we define the $i$-th 
unstable direction 
\begin{equation}
 {f}_i = 
{ {\cal T}  {e}_i- \sum_{j=1}^{i-1}
 ({\cal T}  {e}_i,  {f}_j) {f}_{j}
\over 
| {\cal T}  {e}_i- \sum_{j=1}^{i-1}
 ({\cal T}  {e}_i,  {f}_j) {f}_{j} |
\label{fi}
}.
\end{equation}
Since $\{  {f}_i,\  1  \le i \le 2N \}$ is a set of orthonormal
unit vectors in the tangent space at $\Gamma(t)$, we can find
an orthogonal matrix  ${\cal F}(t,\Gamma(0))$ given by
\begin{equation}
 {f}_i={{\cal F}(t,\Gamma(0))}  {e}_i.
\end{equation}
Further, from Eq. (\ref{fi}), ${\cal T} {e}_i$ is written as
\begin{equation}
{\cal T}(t, \Gamma(0)) {e}_i
=\sum_{k} L_{ik}(t,\Gamma(0)) {\cal F}
(t, \Gamma(0))  {e}_k, 
\label{evol:e}
\end{equation}
where $L_{ij}$ is the $(i,j)$-element of an lower triangle
matrix ${\cal L}$.  Eq. (\ref{evol:e}) is  the Gram-Schmidt 
decomposition of the matrix ${\cal T}$. Since the diagonal element
will be particularly important below, we write it explicitly as
\begin{equation}
L_{ii}=| {\cal T}  {e}_i- \sum_{j=1}^{i-1}
 ({\cal T}  {e}_i,  {f}_j) {f}_{j} |.
\end{equation}

\subsection{Convergence property}

As mentioned above, $ {f}_i$ indicates the $i$-th unstable
direction only when  $t \rightarrow \infty$. 
Let $ {e}_{i*}(\Gamma(t))$ be the 'true' $i$-th 
unstable direction at $\Gamma(t)$. In order to have 
$ {e}_{i*}(\Gamma(t))$  within a certain accuracy, 
we need to confirm
\begin{equation}
d({\cal F}(t, \Gamma(0))  {e}_i,
  {e}_{*i}(\Gamma(t))) \le \epsilon,
\label{conv}
\end{equation}
where $\epsilon $ is a small number related to the accuracy we require,
and $d( {e},  {e'})$ is the absolute value
of  the sine of the  angle between two unit vectors 
$  {e}$ and $ {e'}$.
\begin{equation}
d( {e},  {e'})= \sqrt{1-(e,e')^2}.
\end{equation}
However, since we do not have  $ {e}_{*i}(\Gamma(t))$ yet,
we cannot confirm whether or not  Eq. (\ref{conv}) is satisfied. 
Then, instead of Eq. (\ref{conv}), we check  the condition 
\begin{equation}
d ({\cal F}(t, \Gamma(0))  {e}_i,
   {\cal F}(t, \Gamma(0))  {e'}_i) \le \epsilon
\label{4:conv2}
\end{equation}
for two sets of orthogonal unit vectors 
$\{  {e}_i, 1\le i \le 2N \}$ and
$\{  {e'}_i, 1 \le i \le 2N \}$
which are made randomly.  When Eq. (\ref{4:conv2}) is satisfied,
we assume that the true $i$-th unstable direction is determined by
\begin{equation}
 {e}_{*i}(\Gamma(t)) \simeq {\cal F}(t, \Gamma(0))  {e}_i
\end{equation}
within an accuracy we require.

When we numerically obtain  $ {e}_{*i}(\Gamma(0))$  at an arbitrary
point $\Gamma(0)$,  we consider a trajectory segment 
$\{ \Gamma(t), \ -t_b \le t \le 0 \}$ and check the condition
\begin{equation}
d ({\cal F}(t_b, \Gamma(-t_b))  {e}_i),
   {\cal F}(t_b, \Gamma(-t_b))  {e}'_i)) 
\le \epsilon,
\label{conv21}
\end{equation}
for   sufficiently large   $t_b$. 
When Eq. (\ref{conv21}) is satisfied, we assume 
\begin{equation}
 {e}_{*i}(\Gamma(0)) \simeq {\cal F}(t_b, \Gamma(-t_b)) 
 {e}_i
\end{equation}
within an accuracy we require. 

We do not know a mathematical condition under which 
Eq. (\ref{4:conv2}) is satisfied.  In the argument below, 
we assume that Eq. (\ref{4:conv2}) is satisfied
and that a set of vectors $\{ {e}_{*i}(\Gamma)\}$
is determined for an arbitrary point $\Gamma$.
Once $ {e}_{*i}(\Gamma(0))$ is determined,
$ {e}_{*i}(\Gamma(t))$ is calculated by
\begin{equation}
 {e}_{*i}(\Gamma(t))={\cal F}(t,\Gamma(0)) {e}_{*i}(\Gamma(0)).
\end{equation}

\subsection{Lyapunov vectors}

So far, we have stated that $ {e}_{*i}(\Gamma(t))$ 
indicates the $i$-th unstable direction at $\Gamma(t)$. 
More precisely, 
$ {e}_{*i}(\Gamma(t))$ indicates  the most unstable 
direction in the orthogonal co-space of the subspace spanned 
by  $\{  {e}_{*j}(\Gamma(t)), \ 1 \le j \le i-1 \}$
in the tangent space at $\Gamma(t)$. 
The term {\it orthogonal co-space}  in this statement 
leads to  a non-favorable property of $ {e}_{*i}(\Gamma(t))$:
$ {e}_{*i}(\Gamma(t))$ does {\it not}  satisfy
\begin{equation}
{\cal T}(t,\Gamma(0))  {e}_{*i}(\Gamma(0))  \propto   
 {e}_{*i}(\Gamma(t))
\end{equation}
except for the case $i=1$.  This seems a little bit strange,
because the unstable nature should be defined as something consistent 
along the trajectory. Then, we define a set of vectors
$\{ {\xi}_i(\Gamma(t)), \ 1 \le i \le 2N\}$ which satisfies 
the two conditions. The first condition is the transitivity
\begin{equation}
{\cal T}(t,\Gamma(0))  {\xi}_i(\Gamma(0)) \propto
 {\xi}_i(\Gamma(t)),
\end{equation}
and the second condition is that  the vector space generated 
by $\{  {e}_{*j}, \ 1 \le j \le i\} $ is spanned by 
$\{  {\xi}_j, \ 1 \le j \le i\} $. The second condition is expressed
by 
\begin{equation}
\sum_j A_{ij}  {\xi}_j(\Gamma(t))
= {e}_{*i}(\Gamma(t)),
\label{trans}
\end{equation}
where  $A_{ij}=0$ for $i <j$, and $A_{ij}$ is regarded as 
the $(ij)$-element of  a lower triangle matrix ${\cal A}$.

Now, we  define the $i$-th expansion factor $\Lambda_i(t,\Gamma(0))$
in the $i$-th unstable direction
\begin{equation}
{\cal T}(t,\Gamma(0))  {\xi}_i(\Gamma(0)) =
\Lambda_i(t,\Gamma(0))  {\xi}_i(\Gamma(t)).
\label{Lam}
\end{equation}
We call  $ {\xi}_i$  the $i$-th Lyapunov vector.\cite{GD}
In order to determine uniquely the value of $\Lambda_i$,
we assume the normalization condition that the volume of
the parallelaid made by $\{\xi_j, \ 1 \le j \le i\} $
is the unity. This condition is expressed by
\begin{equation}
|\wedge_{j=1}^i  {\xi}_j|=1
\label{normal}
\end{equation}
for $1 \le i \le 2N$. (See Appendix.) We also assume that $A_{ii}$
is positive. Under these conditions, we can prove  that 
the $i$-th expansion factor $\Lambda_i(t,\Gamma(0))$ is calculated
by the Gram-Schmidt decomposition Eq. (\ref{evol:e}) with 
$ {e}_{*i}(\Gamma(0))$. 

(proof)

{}From Eq. (\ref{evol:e}), we have 
\begin{equation}
{\cal T}(t, \Gamma(0)) {e}_{*i} (\Gamma(0))
=\sum_{k} L_{ik}(t,\Gamma(0)) {\cal F}(t, \Gamma(0)) 
 {e}_{*k}(\Gamma(0)).
\label{evol:es}
\end{equation}
Using Eqs. (\ref{trans}) and (\ref{Lam}), we rewrite the 
left-hand side of Eq. (\ref{evol:es}) as
\begin{eqnarray}
&\phantom{=}&
\sum_j A_{ij}(\Gamma(0)) {\cal T}(t, \Gamma(0))
  {\xi}_j(\Gamma(0)) \\
&=&
\sum_j A_{ij}(\Gamma(0)) \Lambda_j(t, \Gamma(0))
 {\xi}_j(\Gamma(t)) \\
&=&
\sum_{jk} A_{ij}(\Gamma(0)) \Lambda_j(t, \Gamma(0))
({\cal A}(\Gamma(t))^{-1})_{jk}  {e}_{*k}(\Gamma(t)) \\
&=& 
\sum_{jk} A_{ij}(\Gamma(0)) \Lambda_j(t, \Gamma(0))
({\cal A}(\Gamma(t))^{-1})_{jk} {\cal F}(t, \Gamma(0))
 {e}_{*k}(\Gamma(0)).
\label{evol:es2}
\end{eqnarray}
Comparing the right-hand side of Eq. (\ref{evol:es}) and
Eq. (\ref{evol:es2}), we find 
\begin{equation}
L_{ik}(t,\Gamma(0))= \sum_{j} A_{ij}(\Gamma(0)) \Lambda_j(t, \Gamma(0))
({\cal A}(\Gamma(t))^{-1})_{jk}.
\label{starev}
\end{equation}
Further, {}from Eqs. (\ref{trans}) and (\ref{normal}), 
we can easily see
\begin{equation}
|A_{ii}|=1
\label{adiag}
\end{equation}
for $ 1 \le i \le 2N$.  (See Appendix.) Since  $A_{ii}$ is assumed 
to be positive, $A_{ii}=1$. Then,  Eq. (\ref{starev}) yields
\begin{equation}
\Lambda_i(t, \Gamma(0))=L_{ii}(t,\Gamma(0)).
\label{lamstarev}
\end{equation}
In this way, the $i$-th expansion factor can be calculated numerically.

\hfill (q.e.d)

\subsection{Lyapunov exponent}

The $i$-th  expansion ratio $\lambda_i(\Gamma(t))$ at $\Gamma(t)$ 
is defined as
\begin{equation}
\der{\Lambda_i(t,\Gamma(0))}{t}=\lambda_i(\Gamma(t))
\Lambda_i(t,\Gamma(0)).
\end{equation}
The long time average of the $i$-th  expansion ratio 
$\lambda_i(\Gamma(t))$ is  called the $i$-th Lyapunov exponent,
which is given by
\begin{eqnarray}
\bar \lambda_i &=& 
\lim_{\tau\rightarrow \infty}{1\over \tau} \int_0^\tau
 dt \lambda_i(\Gamma(t)),
\\
&=&
\lim_{\tau \rightarrow \infty }{1\over \tau} 
\log \Lambda_i(\tau,\Gamma(0)).
\end{eqnarray}
$\bar \lambda_i$ is sometimes called the 'local' Lyapunov exponent
because $\bar \lambda_i$ depends on $\Gamma(0)$.
However, from the ergodic 
theorem, $\bar \lambda_i$ has a same value for almost all 
$\Gamma(0)$ with respect to the microcanonical measure.
Since we assume the ergodicity of the microcanonical measure,
we do not take care of the local nature of the Lyapunov exponent.

As clearly seen from the method of construction  of Lyapunov 
vectors,  we find 
\begin{equation}
\bar \lambda_1 \ge \bar \lambda_2 \cdots \ge  \bar \lambda_{2N}.
\end{equation}
In Hamiltonian systems, there are at least two zero Lyapunov 
exponents whose Lyapunov vectors indicate the normal direction
of the energy surface and the tangential direction of the 
trajectory.  
In the argument below, we assume that there are $ N_p$ 
positive Lyapunov exponents. Unless the system has a further 
conservation law such as momentum conservation, $N_p$ equals to $N-1$.

The {\it information loss rate} at $\Gamma(t)$, $h(\Gamma(t))$,
is defined as the sum of the expansion ratios with the positive Lyapunov 
exponents. 
\begin{eqnarray}
h(\Gamma(t)) &=& \sum_{\bar \lambda_i > 0} \lambda_i(\Gamma(t)),\\
      &=& \sum_{i=1}^{N_p} \lambda_i(\Gamma(t)).
\end{eqnarray}
Notice that $h(\Gamma(t))$ represents the volume expansion 
ratio of the $N_p$-dimensional unstable space.
That is,  $h(\Gamma(t))$ is rewritten as 
\begin{equation}
h(\Gamma(t))={d \over dt} 
\log |\wedge_{i=1}^{N-1}  {\cal T}(t,\Gamma(0))
 {\xi}_{i}(\Gamma(0))|.
\label{4:37}
\end{equation}
The long time average of the information loss rate, $\bar h$,
has the same value for almost all initial conditions  with respect 
to the microcanonical  measure. It has been known that 
$\bar h$ is identical to the Kolmogorov-Sinai entropy
when the system is hyperbolic.\cite{Pesin}

\subsection{Contraction ratio}

Let us recall that the expansion factor $\Lambda_i$
is calculated by the Gram-Schmidt decomposition under 
the normalization condition Eq. (\ref{normal}). 
However, this normalization lacks the balance between the 
unstable and stable directions. 
Since Hamiltonian systems possess a time reverse symmetry, 
such unbalance will cause theoretical complicatedness.
In order to recover the symmetry, we introduce  a new set of vectors
$\{  {\xi}_i^{(s)},\ 1 \le i \le 2N \} $ given by
\begin{equation}
 {\xi}_i^{(s)} = c_i {\xi}_{2N-i+1},
\label{etac}
\end{equation}
where $c_i$ is a positive number determined so as to satisfy
\begin{equation}
|\wedge_{j=1}^{i}  {\xi}_j^{(s)}|=1
\label{normeta}
\end{equation}
for  $1\le i \le 2N$. The set of vectors 
$\{ {\xi}_i^{(s)}, \ 1 \le i \le 2N \}$
is made to respect stable directions from the most stable one.

We now define the contraction factor $\Lambda_i^{(s)}$ and
contraction ratio $\lambda_i^{(s)}$ as
\begin{eqnarray}
{\cal T}(t,\Gamma(0))  {\xi}_i^{(s)}(\Gamma(0)) 
&=& {1 \over \Lambda_i^{(s)}(t,\Gamma(0))} 
 {\xi}_i^{(s)}(\Gamma(t)), 
\label{cf} \\
\lambda_i^{(s)}(\Gamma(t)) &=&
{d \over dt} \log  \Lambda_i^{(s)}(t,\Gamma(0)).
\end{eqnarray}
As will be seen in the next section, the  contraction ratio
is related to the  expansion ratio of the time-reversed 
trajectory and this relation plays a role in  simplifying  arguments.
In particular, the following estimation will be  utilized.
\begin{equation}
\sum_{i=1}^{N_p} \lambda_{2N+1-i} 
=-\sum_{i=1}^{N_p}\lambda_i^{(s)}+{d \over dt}o(N).
\label{4:conj}
\end{equation}
Here, the last term represents the time derivative 
of a function whose value is much smaller than $N$
when $N \rightarrow \infty$. Note that the left-hand side
and the first term of the right-hand side are the order of $N$.

(proof)

Substituting Eq. (\ref{etac}) intro Eq. (\ref{cf}), we have
\begin{equation}
c_i(\Gamma(0)) \Lambda_{2N+1-i}(t,\Gamma(0)) 
={1 \over \Lambda_i^{(s)}(t,\Gamma(0))} c_i(\Gamma(t)).
\end{equation}
The time derivative of the logarithm of the both-hand sides
yields
\begin{equation}
\lambda_{2N+1-i}(\Gamma(t)) 
= -\lambda_i^{(s)}(\Gamma(t))+{d \over dt} 
\log \left( {c_i(\Gamma(t)) \over c_i(\Gamma(0))} \right) .
\label{lams}
\end{equation}
We thus obtain
\begin{equation}
\sum_{i=1}^{N_p} \lambda_{2N+1-i}(\Gamma(t)) 
= - \sum_{i=1}^{N_p}
\lambda_i^{(s)}(\Gamma(t))
+{d \over dt} \sum_{i=1}^{N_p}\log 
\left( {c_i(\Gamma(t)) \over c_i(\Gamma(0))} \right).
\label{lams2}
\end{equation}

Let us evaluate the second term of the right-hand side of 
Eq. (\ref{lams2}).  We first define  an 'angle' $\phi_i$ as
\begin{equation}
|\wedge_{j=1}^{2N}  {\xi}_j| = 
|\wedge_{j=1}^{2N-i}  {\xi}_j|
|\wedge_{j=2N+1-i}^{2N}  {\xi}_j|
 \sin \phi_i,
\label{angle}
\end{equation}
where $0 \le \phi_i \le \pi/2$. (See Appendix.)  
By using the normalization condition of $\{\xi_i \}$,
we rewrite Eq. (\ref{angle}) as
\begin{equation}
|\wedge_{j=2N+1-i}^{2N}  {\xi}_j| \sin \phi_i=1.
\label{angle2}
\end{equation}
Using Eqs. (\ref{etac}) and (\ref{normeta}), we obtain
\begin{equation}
c_1\cdots c_i =\sin \phi_i.
\label{angle3}
\end{equation}
This leads to
\begin{equation}
\sum_{i=1}^{N_p}\log 
\left( {c_i(\Gamma(t)) \over c_i(\Gamma(0))} \right)
=\log\left( {\sin \phi_{N_p}(\Gamma(t))\over \sin \phi_{N_p}(\Gamma(0))
 }\right) =o(N),
\end{equation}
where we have assumed 
\begin{equation}
\sin(\phi_{N_p}(\Gamma)) \gg O( \exp(-N) ),
\end{equation}
which may be ensured by the condition that the
unstable and stable manifolds intersect transversally.

\hfill (q.e.d.)

\subsection{Weight on trajectory segments}

We consider a weight on the trajectory segment 
$\{\Gamma(t),\ 0 \le t \le \tau \}$.  The weight, 
$W(\{\Gamma(t),\ 0 \le t \le \tau \})$, is  a conditional
probability finding trajectory segments remaining in a small
tube around  $\{\Gamma(t),\ 0 \le t \le \tau \}$ when the initial 
condition is chosen in a small region around $\Gamma(0)$. 
More explicitly, the weight $W$ is defined in the following way.

Suppose that the phase space is decomposed into small cells 
$\{ \Delta_j \}$ with a sufficiently small size $\epsilon$
and that $\Gamma(0)$ is included in the $i$-th cell $\Delta_i$.
Then, the number of cells which intersect with  $U_\tau(\Delta_i)$,
$N(\tau,\epsilon)$, can be counted.  
We can choose the value of $\epsilon$  so that the region 
$U_\tau(\Delta_i)$  remains in a linear regime around 
$U_\tau(\Gamma(0))$. This condition may be given by
\begin{equation}
\epsilon  \ll \epsilon_c(\tau),
\label{pcon}
\end{equation}
where the value of $\epsilon_c(\tau)$ is determined 
by nonlinear properties of dynamical systems.  
Under this condition, $N(\tau,\epsilon)$
measures the number of distinguishable 
trajectory segments 
starting from the neighborhood of $\Gamma(0)$. 
Therefore, we define  the weight as
\begin{equation}
W(\{\Gamma(t),\ 0 \le t \le \tau \})={1 \over N(\tau,\epsilon)}.
\label{w:def}
\end{equation}

Then, we can show 
\begin{equation}
W(\{\Gamma(t),\ 0 \le t \le \tau\})=
|\wedge_{i=1}^{N_p}  {\cal T}(-\tau,\Gamma(\tau))
   {\xi}_i(\Gamma(\tau))|
\label{weight}
\end{equation}
in an appropriate limit of large $\tau$ and small $\epsilon$.
(Since $ \epsilon_c(\tau)\rightarrow 0 $ for the limit 
 $\tau \rightarrow \infty$, we need to take care of a delicate problem 
 of double limits. Nevertheless, we assume simply that we can
choose an appropriate  asymptotic limit.)

(proof)

Consider the time evolution of a small region $\Delta_i$.
The region expands and contracts in the unstable and stable directions, 
respectively. After a sufficiently long time,  
the region almost collapses into 
the $N_p$-dimensional unstable manifold, and intersects with cells
in the unstable directions. Since  the $N_p$-dimensional volume 
element in the unstable manifold at $\Gamma(\tau)$ is written as
$ \wedge_{i=1}^{N_p}   {\xi}_i(\Gamma(\tau))$, 
we expect
\begin{eqnarray}
N(\tau,\epsilon) &=& 
{
|\wedge_{i=1}^{N_p}   {\xi}_i(\Gamma(\tau))|
\over
|\wedge_{i=1}^{N_p}  {\cal T}(-\tau, \Gamma(\tau))
 {\xi}_i(\Gamma(\tau))| } \\
&=&
|\wedge_{i=1}^{N_p}  {\cal T}(-\tau, \Gamma(\tau))
 {\xi}_i(\Gamma(\tau))|^{-1}
\end{eqnarray}
in an appropriate limit of large $\tau$ and small $\epsilon$.
Substituting this into Eq. (\ref{w:def}) leads to 
Eq. (\ref{weight}). 

\hfill (q.e.d)

Further, by the relation Eqs. (\ref{4:37}),
Eq. (\ref{weight}) becomes a simpler form 
\begin{equation}
W(\{\Gamma(t),\ 0 \le t \le \tau\})=\exp(-\int_0^\tau dt h(\Gamma(t))).
\label{weight2}
\end{equation}

\subsection{Time dependent case}

To this point in this section, we have assumed that the value of $\alpha$
is not changed in time. In  this subsection, we  briefly discuss
Lyapunov analysis for systems with a time dependent parameter.
When the value of $\alpha$ depends on $t$, the evolution map from 
$t=t_0$ to $t=t_1$ depends on the absolute time $t_0$ and $t_1$.
Therefore, it takes the form $U_{t_1,t_0}$, and the linearized evolution 
map is written as ${\cal T}(t_1,t_0; \Gamma(t_0))$. 

The Lyapunov analysis in such a case may be reconsidered carefully. 
However, we do not need general arguments. In the systems in question,
the value of $\alpha$ is changed  during a finite time 
interval $[\tau_i,\tau_f]$, where $ 0 \ll \tau_i \le  \tau_f \ll \tau$.
Therefore, for example, 
the $i$-th Lyapunov vectors at $\Gamma(0)$ and $\Gamma(\tau)$ 
can be defined as $\xi_i(\Gamma(0))$ and $\xi_i(\Gamma(\tau))$, 
respectively. 

Although the expansion factors, Lyapunov exponents and  information 
loss rate do not make a sense in general, the argument on the weight 
$W$ is still valid. We can write 
\begin{equation}
W(\{\Gamma(t),\ 0 \le t \le \tau\})=
|\wedge_{i=1}^{N_p}  {\cal T}(0,\tau;\Gamma(\tau))
   {\xi}_i(\Gamma(\tau))|.
\label{weight:g}
\end{equation}
for sufficiently large $\tau$ and small $\epsilon$. 
We also define the {\it actual information loss rate} 
as the generalization of  Eq. (\ref{4:37})
\begin{equation}
h_{\rm a}(t,\Gamma(0))={d \over dt} 
\log |\wedge_{i=1}^{N-1}  {\cal T}(t,0;\Gamma(0))
 {\xi}_{i}(\Gamma(0))|.
\label{ainf}
\end{equation}

\subsection{Liouville's theorem}

In this subsection, we review a proof of the Liouville's theorem
which states that the $2N$-dimensional volume element keeps its volume 
along the trajectory. It is important to understand that the Liouville's 
theorem holds even when  the value of $\alpha$ is changed in time.

(proof)

We have  the Hamiltonian equation
\begin{equation}
\der{\Gamma(t)}{t} = -{\cal J}
\left.\pder{H(\Gamma, \alpha(t))}{\Gamma}
\right\vert_{\Gamma=\Gamma(t)}.
\end{equation}
(See Eq. (\ref{bigeq}) in Section 2.)
Since  the linearized evolution equation is written as 
\begin{equation}
\der{\delta \Gamma(t)}{t} = -{\cal J}
\left. 
{\partial^2 H(\Gamma, \alpha(t)) \over \partial \Gamma \partial \Gamma}
\right\vert_{\Gamma=\Gamma(t)} \delta \Gamma(t),
\end{equation}
the linearized evolution map ${\cal T}(t,0; \Gamma(0))$
satisfies the equation
\begin{equation}
\der{{\cal T}}{t}=-{\cal J}{\cal B}{\cal T},
\end{equation}
where ${\cal B}$ is a symmetric matrix.
We then  obtain
\begin{eqnarray}
\der{({\cal T}^\dagger {\cal J} {\cal T})}{t} 
&=&
\der{{\cal T}^\dagger}{t} {\cal J} {\cal T} +
{\cal T}^\dagger {\cal J } \der{{\cal T}}{t} \\
&=& 
-({{\cal T}^\dagger} {\cal B}{ \cal J}^\dagger {\cal J }{\cal T} +
{\cal T}^\dagger {\cal J }{ \cal J }{\cal B}{\cal  T}) \\
&=& 0
\label{4:83},
\end{eqnarray}
where we have used the equality
\begin{equation}
{\cal J}^\dagger{\cal J}=-{\cal J}{\cal J}=1.
\end{equation}
Since ${\cal T}(0,0;\Gamma(0))= 1$, Eq. (\ref{4:83}) leads to 
\begin{equation}
{\cal T}^\dagger {\cal J}{\cal T}={\cal J}.
\label{symp}
\end{equation}
The determinant of the both-hand sides gives 
\begin{equation}
{\rm det } [{\cal T}^\dagger {\cal T}] = 1.
\end{equation}

Let $ \{e_i, \  1\le i \le 2N \}$ be an orthogonal set of 
unit vectors defined in the tangent space at $\Gamma(0)$.
The time evolution of the $2N$ dimensional volume element
$\wedge_{i=1}^{2N} e_i $ is given by 
$\wedge_{i=1}^{2N} {\cal T} e_i $, and  its volume is calculated as
\begin{eqnarray}
|\wedge_{i=1}^{2N} {\cal T} e_i| &=& 
\sqrt{ {\rm det} {\cal T}{\cal T}^\dagger}  \\
                                 &=& 1.
\end{eqnarray}
(See Appendix.)
Therefore, the $2N$-dimensional volume element keeps its volume 
along the trajectory. 
\hfill (q.e.d.)

Further, using the Liouville's theorem, we can prove that the equality
\begin{equation}
\sum_{i=1}^{2N}\lambda_i(\Gamma(t))=0
\label{zerosum}
\end{equation}
holds when the value of $\alpha$ is not changed in time. 

(proof)

Since the the value of $\alpha$ is not changed in time, we obtain
\begin{eqnarray}
\sum_{i=1}^{2N}\lambda_i(\Gamma(t)) 
&=& 
{d\over dt} \log \sum_{i=1}^{2N}\Lambda_i(t,\Gamma(0)) \\
&=& 
{d\over dt}\log |\wedge_{i=1}^{2N} {\cal T}(t,\Gamma(0)) \xi_i(\Gamma(0))| \\
&=& 0
\end{eqnarray}
where the Liouville's theorem is used to obtain the last line.

\hfill (q.e.d)

\section{Reversibility}

\subsection{Reversibility in time evolution}

We  define a matrix $\rev $ as
\begin{eqnarray}
(\rev \Gamma)_i  &=& q_i, \\
(\rev \Gamma)_{i+N}  &=& -p_i, 
\end{eqnarray}
where  $1 \le i \le N$. The matrix $\rev$ corresponds to the time reverse
operator acting on the phase space point. We assume that 
the Hamiltonian under consideration possesses the time reverse
symmetry  
\begin{equation}
H(\rev \Gamma,\alpha)=H(\Gamma,\alpha).
\label{symH}
\end{equation}

Let ${  U}_{t,0}$ and $\tilde {   U}_{t,0}$  be the evolution maps
for Hamiltonian equations with $\alpha()$ and $\tilde \alpha()$,
respectively, where  we have defined the time-reversed protocol
of the parameter change $\tilde \alpha()$ as
\begin{equation}
\tilde \alpha(t)=\alpha(-t).
\end{equation}
Then, owing to the symmetry property Eq. (\ref{symH}),  the identity
\begin{equation}
{  U}_{t,0}=\rev \tilde {   U}_{-t,0}  \rev.
\label{revident}
\end{equation}
holds. 

(proof)

Let  $\{\Gamma(t) \}$ and $\{\tilde \Gamma(t) \}$  be  
trajectories  given  by 
\begin{eqnarray}
\Gamma(t) &=& {  U}_{t,0}(\Gamma(0)), 
\label{ut}\\ 
\tilde \Gamma(t) &=& \tilde {  U}_{t,0}(\tilde \Gamma(0)), 
\label{utt}
\end{eqnarray}
where  $\Gamma(0)$ and $\tilde \Gamma(0)$ are the initial conditions
which satisfy the relation 
\begin{equation}
\tilde \Gamma(0)=\rev \Gamma(0).
\label{trevini}
\end{equation}
{}From Eq. (\ref{utt}), we obtain
\begin{eqnarray}
-\der{\tilde \Gamma(-t)}{t} &=& 
-{\cal J} \left.\pder{H(\Gamma, \tilde \alpha(-t))}{\Gamma}
\right\vert_{\Gamma=\tilde \Gamma(-t)},\\
&=& 
-{\cal J}\left.\pder{H(\Gamma, \alpha(t))}{\Gamma}
\right
\vert_{\Gamma=\tilde \Gamma(-t)},
\label{5-3}
\end{eqnarray}
where we have used the equation of motion in the form 
Eq. (\ref{bigeq}) with the matrix ${\cal J}$ satisfying Eq. (\ref{bigJ}).  
On the other hand, Eq. (\ref{ut}) leads to
\begin{eqnarray}
\rev\der{\Gamma(t)}{t}
&=& -\rev {\cal J}
\left.\pder{H(\Gamma,\alpha(t))}{\Gamma}
\right\vert_{\Gamma=\Gamma(t)}, \\
&=&  {\cal J} \rev 
\left.\pder{H(\Gamma,\alpha(t))}{\Gamma}
\right\vert_{\Gamma=\Gamma(t)}, \\
&=&  {\cal J} 
\left.\pder{H(\Gamma,\alpha(t))}{(\rev \Gamma)}
\right\vert_{\Gamma=\Gamma(t)},\\
&=&  {\cal J} 
\left.\pder{H(\rev\Gamma,\alpha(t))}{(\rev \Gamma)}
\right\vert_{\Gamma=\Gamma(t)}
,\\
&=&  {\cal J} 
\left.\pder{H(\Gamma,\alpha(t))}{\Gamma}
\right\vert_{\Gamma=\rev\Gamma(t)}.
\label{5-4}
\end{eqnarray}
Here, the second line is obtained by the relation 
\begin{equation}
\rev {\cal J}+ {\cal J}\rev=0,
\end{equation}
the third line is derived from the relation $\rev\rev=1$, 
and the equality 
of the forth line comes from the symmetry property Eq. (\ref{symH}).

Comparing Eqs. (\ref{5-3}) and (\ref{5-4}), we find that
$\tilde \Gamma(-t)$  and $\rev \Gamma(t) $ obey the
same evolution equation.  Recalling the relation for the initial 
conditions Eq. (\ref{trevini}), we conclude 
\begin{equation}
\tilde \Gamma(-t)=\rev \Gamma(t).
\label{eq5-5}
\end{equation}
By using Eqs. (\ref{ut}) and  (\ref{utt}), Eq. (\ref{eq5-5}) is rewritten as
\begin{equation}
\tilde {  U}_{-t,0}(\rev \Gamma(0))=\rev {  U}_{t,0}(\Gamma(0)).
\end{equation}
Since $\Gamma(0)$ is arbitrary, Eq. (\ref{revident}) holds.

\hfill (q.e.d)

In the argument below, $\tilde \Gamma(t) $ will be assumed to be
given  by Eq. (\ref{eq5-5}).

\subsection{Reversibility in Lyapunov analysis}

First, from Eq. (\ref{revident}), we have
\begin{equation}
{\cal T}(t,0; \Gamma(0))=\rev \tilde {\cal T}(-t,0;\tilde\Gamma(0))
\rev ,
\label{revidentt}
\end{equation}
where ${\cal T}(t,0;\Gamma(0))$ and 
$\tilde {\cal  T}(t,0;\tilde \Gamma(0))$ 
are the linearized evolution maps around the 
trajectories $\Gamma()$ and $\tilde \Gamma()$.
In particular, 
when the value of $\alpha$ is not changed, the equality 
$\tilde {\cal T}={\cal T}$ holds. We then  prove the identities
\begin{eqnarray}
{\xi}_i( \Gamma)&=& \rev {\xi}^{(s)}_{i}(\rev \Gamma), 
\label{vecsym} \\
{\lambda}_i(\Gamma)&=& {\lambda}_{i}^{(s)}(\rev \Gamma),
\label{chisym} \\
h(\Gamma(t))- h(\tilde \Gamma(-t))
&=& {d \over dt} o(N).
\label{conv0}
\end{eqnarray}

(proof) 

Using the matrices defined as
\begin{eqnarray}
{\cal X}_{ij}(\Gamma) &=& ({\xi}_j(\Gamma) )_i,\\
{\cal X}^{(s)}_{ij}(\Gamma) &=& ({\xi_j^{(s)}}(\Gamma) )_i,
\end{eqnarray}
we can write ${\cal T}(t,\Gamma(0))$ in the two forms
\begin{eqnarray}
{\cal T}(t,\Gamma(0))&=&{\cal X}(\Gamma(t)){\cal M}(t,\Gamma(0))
{\cal X}(\Gamma(0))^{-1},
\label{matherm} \\
{\cal T}(t,\Gamma(0))&=&{\cal X}^{(s)}(\Gamma(t))
{\cal M}^{(s)}(t,\Gamma(0)){\cal X}^{(s)}(\Gamma(0))^{-1},
\label{matherms}
\end{eqnarray}
where ${\cal M}(t,\Gamma(0))$ and ${\cal M}^{(s)}(t,\Gamma(0))$ 
are diagonal matrices whose $(i,i)$-elements are given by 
$\Lambda_i(t,\Gamma(0))$ and $\Lambda^{(s)}_i(t,\Gamma(0))^{-1}$,
respectively. Using Eq. (\ref{matherms}), we rewrite the right-hand
side of Eq. (\ref{revidentt}) as
\begin{eqnarray}
&\phantom{=}&  
\rev {\cal X}^{(s)}(\tilde \Gamma(-t)){\cal M}^{(s)}(-t,\tilde \Gamma(0))
{\cal X}^{(s)}(\tilde \Gamma(0))^{-1} \rev,\\
&=&
\rev {\cal X}^{(s)}(\rev  \Gamma(t)){\cal M}^{(s)}(-t,\tilde \Gamma(0))
{\cal X}^{(s)}(\rev \Gamma(0))^{-1}  \rev,
\label{528}
\end{eqnarray}
where we have used $\tilde {\cal T}={\cal T}$. 
Comparing Eq. (\ref{528}) with the right-hand side of Eq. (\ref{matherm}),
we obtain 
\begin{eqnarray}
 {\cal X}(\Gamma) &=& \rev {\cal X}^{(s)}(\rev \Gamma), 
\label{tildeX} \\
 {\cal M}(t, \Gamma(0))&=& {\cal M}^{(s)}(-t, \rev \Gamma(0))
\label{tildeM},
\end{eqnarray}
where notice the normalization conditions of $\xi_i$ and 
$\xi^{(s)}_i$ given by Eqs. (\ref{normal}) and (\ref{normeta}). 

Equation (\ref{tildeX}) is equivalent to Eq. (\ref{vecsym}),
and  Eq. (\ref{tildeM}) leads to Eq. (\ref{chisym}),
because  of  the equality
\begin{eqnarray}
{d\over dt}\log\Lambda_i^{(s)}(-t,\rev \Gamma(0))^{-1}
&=& \lambda^{(s)}_i(\tilde \Gamma(-t)),\\
&=& \lambda^{(s)}_i(\rev \Gamma(t)).
\end{eqnarray}

Furthermore,  $h(\Gamma(t))$ is expressed in terms of $\{ \lambda_i^{(s)} \}$
in the following way.
\begin{eqnarray}
h(\Gamma(t)) &=& \sum_{i=1}^{N_p} \lambda_i(\Gamma(t)), \\
      &=& -\sum_{i=1}^{N_p} \lambda_{2N+i-1}(\Gamma(t)),  \\
      &=& \sum_{i=1}^{N_p} \lambda_i^{(s)}(\Gamma(t))
          +{d \over dt}o(N), 
\label{5p-1}
\end{eqnarray}
where the second and third lines come from Eqs.  (\ref{zerosum}) 
and  Eq. (\ref{4:conj}), respectively. 
On the other hand,  from the symmetry property Eq. (\ref{chisym}), 
$h(\tilde \Gamma(-t))$ is written as
\begin{eqnarray}
h(\tilde \Gamma(-t))
&=& \sum_{i=1}^{N_p} \lambda_i(\tilde \Gamma(-t)), \\
&=& \sum_{i=1}^{N_p} \lambda_i(\rev  \Gamma(t)), \\
&=& \sum_{i=1}^{N_p} \lambda_i^{(s)} (\Gamma(t)).
\label{5p-2}
\end{eqnarray}
Comparing Eqs. (\ref{5p-1}) and (\ref{5p-2}), we obtain
Eq.(\ref{conv0}). 

\hfill (q.e.d.)

Using these identities, we can express the weight on the
trajectory segment by 
the actual information loss of the time reversed trajectory
\begin{equation}
W(\{\Gamma(t), \ 0 \le t \le \tau\})= 
\int_0^\tau dt \tilde h_a(-t,\tilde \Gamma(-\tau))\exp(o(N)),
\label{wha1}
\end{equation}
where  the actual information loss rate along the time-reversed 
trajectory $\tilde h_a$ is defined as 
\begin{equation}
\tilde h_a(t,\tilde \Gamma(-\tau))={d \over dt} 
\log
|\wedge_{i=1}^{N-1} \tilde {\cal T}(t,-\tau;\Gamma(-\tau))
  {\xi}_{i}(\tilde \Gamma(-\tau))|.
\end{equation}

(prove)

Let us recall the expression of the weight Eq. (\ref{weight:g}).
\begin{equation}
W(\{\Gamma(t), \ 0 \le t \le \tau\})  = 
|\wedge_{i=1}^{N_p}  {\cal T}(0,\tau;\Gamma(\tau))
  {\xi}_i(\Gamma(\tau))| . 
\end{equation}
The right-hand side is rewritten in the following way
\begin{eqnarray}
&\phantom{=}&
|\wedge_{i=1}^{N_p} \rev \tilde {\cal T}(0,-\tau; \tilde \Gamma(-\tau))
\rev \rev  {\xi}_i^{(s)}(\rev \Gamma(\tau))| \\
&=&
|\wedge_{i=1}^{N_p} \tilde {\cal T}(0,-\tau;\tilde \Gamma(-\tau))
  {\xi}_i^{(s)}(\tilde \Gamma(-\tau))| \\
&=&
|\wedge_{i=1}^{N_p} \tilde {\cal T}(0,-\tau;\tilde \Gamma(-\tau))
  {\xi}_{2N+1-i}(\tilde \Gamma(-\tau))|\exp(o(N)) \\
&=&
|\wedge_{i=1}^{N_p} \tilde {\cal T}(0,-\tau;\tilde \Gamma(-\tau))
  {\xi}_{i}(\tilde \Gamma(-\tau))|\exp(o(N)), \\
&=&
\int_0^\tau dt \tilde h_a(-t,\tilde \Gamma(-\tau))\exp(o(N))
\end{eqnarray}
where the third line is obtained by using an  argument in
section 4.5, and the forth line comes from the Liouville's
theorem.  

\hfill (q.e.d)

Similarly, the weight on the time reversed trajectory segment 
$W(\{\tilde \Gamma(t), \ -\tau \le t \le 0\})$ is written as
\begin{equation}
W(\{\tilde \Gamma(t), \ -\tau \le t \le 0 \})= 
\int_0^\tau dt  h_a(t, \Gamma(0))\exp(o(N)).
\label{wha2}
\end{equation}
(See Eq. (\ref{ainf}) for the definition of $h_a(t, \Gamma(0))$.)

\subsection{Reversibility paradox}

Suppose that there is a trajectory segment 
$\{\Gamma(t),\ 0 \le t \le \tau \}$ from an energy surface $\Sigma_0$
to $\Sigma_1$. Then, the time reversed one
$\{\tilde \Gamma(t),\ -\tau \le t \le 0 \}$ goes from $\Sigma_1$
to $\Sigma_0$. One may wonder how this fact is compatible with
thermodynamic irreversibility.  The essentially same question
was proposed  by Roschmidt, and has been known as the reversibility 
paradox.\cite{Ehren}  A standard answer may be replying on the operational
impossibility of the time reverse operation $\Gamma \rightarrow 
\rev \Gamma$.  If we were allowed to operate the system 
by using the result of the observation of the trajectory, we could 
perform the time reverse operation. This consideration is
related to the Maxwell's demon's problem.\cite{demon} 
However,  the time dependence of $\alpha$ is given 
without any references of trajectories. Thus, in our problem, 
the time reverse operation cannot be realized by  $\alpha()$
and the Maxwell's demon problem does not appear.

However, still the paradox is not resolved completely. 
In order to be compatible with thermodynamic irreversibility, 
there should be asymmetry 
between the trajectory segment $\{\Gamma(t),\ 0 \le t \le \tau \}$ 
and  the time-reversed one 
$\{\tilde \Gamma(t),\ -\tau \le t \le 0 \}$. The asymmetry 
cannot come from a purely mechanical consideration. We must
consider the measure for the ensemble of the initial conditions 
of the time-reversed trajectory segment 
$\{\tilde \Gamma(t),\ -\tau \le t \le 0 \}$.
This ensemble, $\Upsilon_\tau$, is defined as a $(2N-1)$-dimensional set
which satisfies 
\begin{equation}
\tilde U_{0,-\tau}(\Upsilon_\tau)=\Sigma_0.
\label{revUp}
\end{equation}
{}From the reversibility relation Eq. (\ref{revident}),
we obtain
\begin{equation}
\Upsilon_\tau=\rev U_{\tau,0}(\Sigma_0).
\label{Updef}
\end{equation}
Owing to the chaotic nature, $\Upsilon_\tau$ becomes a quite 
complicated set as $\tau$ is large. Here, we describe the 
set $\Upsilon_\tau$ informally.  We focus on the thermodynamic limit
so that the structure of $\Upsilon_\tau$ is clearly seen.

Suppose that the most probable processes $\Sigma_0 \fwdp \Sigma_1$ 
and $\tilde \Sigma_1 \fwdp \Sigma_0$ are realized by the protocols 
of the parameter change $\alpha()$ and $\tilde \alpha()$, respectively. 
Then, from Eq. (\ref{revUp}), $\Upsilon_\tau \cap \tilde \Sigma_1$ 
becomes dominant in $\tilde \Sigma_1$ with respect to
the microcanonical measure for $\tilde \Sigma_1$. On the other hand, 
from Eq. (\ref{Updef}), $\Upsilon_\tau \cap \Sigma_1$ becomes 
dominant in $\Upsilon_\tau$  with respect to the 
microcanonical measure for $\Sigma_0$.  One may wonder that these 
two statements  are apparently contradictory. However, we should note 
that the measures are different when we observe the set $\Upsilon_\tau$.  
Here, it is worthwhile noting that $\Upsilon_\tau \cap \Sigma_1$ 
is {\it not} dominant in $\Sigma_1$ with respect to 
the microcanonical measure for $\Sigma_1$, when $\Sigma_1 \not=
\tilde \Sigma_1$. Therefore, we can imagine that
the set $\Upsilon_\tau$ has a fine structure in energy surfaces
apart from $\tilde \Sigma_1$. In order to represent this heterogeneity
quantitatively, we define a measure 
$\tilde \mu$ for the set $\Upsilon_\tau$ as
\begin{equation}
\tilde \mu(\Delta_\epsilon(\tilde\Gamma(-\tau)); \Upsilon_\tau)
=\lim_{\delta E \rightarrow 0}
{
\mu_{\rm L}( \Delta_\epsilon(\tilde \Gamma(-\tau))
\cap \rev U_{\tau,0}(\Sigma_0\circ \delta E)) 
\over
\mu_{\rm L}(\rev U_{\tau,0}(\Sigma_0 \circ \delta E))},
\end{equation}
where $\mu_{\rm L}$ denotes the $2N$-dimensional Lebesgue measure,
$\Delta_\epsilon(\tilde\Gamma(-\tau))$ is a small region 
with a size $\epsilon$ including 
$\tilde\Gamma(-\tau)$, and $\Sigma_0 \circ \delta E$ represents 
a set of energy  surfaces with width $\delta E$. (See section 2.4.)
We then expect that $\tilde \mu(\Delta_\epsilon(\tilde\Gamma(-\tau)); 
\Upsilon_\tau)$ for  $\tilde\Gamma(-\tau) \in \tilde \Sigma_1$ 
is much larger than $\tilde \mu(\Delta_\epsilon(\tilde\Gamma(-\tau); 
\Upsilon_\tau)$ for $\tilde\Gamma(-\tau) \in  \Sigma_1$.

\subsection{Reversibility in probability}

In spite of the asymmetry between the two sets $\Sigma_0$ and 
$\Upsilon_\tau$, from the reversibility of the time evolution, there 
is a one-to-one correspondence between a set of trajectory 
segments from $\Sigma_0$ to $\rev \Upsilon_\tau$ and a set of 
time-reversed trajectory segments from  $\Upsilon_\tau$ to $\Sigma_0$. 

We now derive the reversibility relation coming from this 
correspondence. In Section 4.7, we  discussed the weight 
on trajectory segments. The weight 
$W(\{\Gamma(t),\ 0 \le t \le \tau\})$ is  a conditional probability  
finding trajectory segments remaining in a small tube around 
$\{\Gamma(t),\ 0 \le t \le \tau\}$ when the initial condition 
is chosen in a small region $\Delta_\epsilon(\Gamma(0))$ 
around $\Gamma(0)$. Then, 
\begin{equation}
 \mu_{\rm mc}(\Delta_\epsilon(\Gamma(0));\Sigma_0) 
W(\{\Gamma(t),\ 0 \le t \le \tau\}) 
\end{equation}
is a probability finding a trajectory segment in a small tube around 
$\{\Gamma(t),\ 0 \le t \le \tau\}$ of all trajectory segments 
from $\Sigma_0$ to $\rev \Upsilon_\tau$. 
The probability, from  the one-to-one  correspondence mentioned above, 
should be equal to a  probability finding a trajectory segment in a 
small tube around $\{\tilde \Gamma(t),\ -\tau \le t \le 0\}$ of 
all trajectory segments from $\Upsilon_\tau$ to $\Sigma_0$.
The latter probability is written as
\begin{equation}
\tilde \mu(\Delta_\epsilon(\tilde\Gamma(-\tau)); \Upsilon_\tau)
W(\{\tilde \Gamma(t),\ -\tau \le t \le 0\}).
\end{equation}
Therefore, we obtain the relation
\begin{equation}
\mu_{\rm mc}(\Delta_\epsilon(\Gamma(0));\Sigma_0) 
W(\{\Gamma(t),\ 0 \le t \le \tau\}) 
=\tilde \mu(\Delta_\epsilon(\tilde\Gamma(-\tau)); \Upsilon_\tau)
W(\{\tilde \Gamma(t),\ -\tau \le t \le 0\}).
\label{mrev}
\end{equation}
This relation will lead to an important equality
related to thermodynamic irreversibility.

\section{Irreversible information loss}

\subsection{Definition}

Let us define the {\it irreversible information loss} $I$ as
\begin{equation}
I(\tau,\Gamma(0))= \log
{ W(\{\tilde \Gamma(t),\ -\tau \le t \le 0\}) \over 
  W(\{\Gamma(t),\ 0 \le t \le \tau\})},
\end{equation}
where the right-hand side depends on $(\tau, \Gamma(0) )$
because the trajectory is given by a solution to the 
deterministic evolution equation. 
Using Eqs. (\ref{wha1}) and (\ref{wha2}), we   can write $I$ as 
\begin{equation}
I(\tau,\Gamma(0)) = \int_0^{\tau}  dt
[h_a(t,\Gamma(0))-\tilde h_a(-t,\tilde \Gamma(-\tau))]+o(N).
\end{equation}
The  expression may represent the meaning  of the term 
'irreversible information loss'. 
Then, from the relation Eq. (\ref{mrev}), we obtain
\begin{equation}
\int_{\Upsilon_\tau} \tilde 
\mu(\Delta_\epsilon(\Gamma(-\tau));\Upsilon_\tau)
=\int_{\Sigma_0}  \mu_{\rm mc}(\Delta_\epsilon (\Gamma(0))) \exp(-I).
\end{equation}
By the normalization condition of the probability,
we have the equality 
\begin{equation} 
\bra \exp(-I) \ket_0=1,
\label{jarz}
\end{equation}
where $\bra \ \ket_0$ denotes the average by 
$\mu_{\rm mc}(d\Gamma;\Sigma_0)$. Using the Jensen's inequality 
\begin{equation} 
\bra \exp(-I) \ket_0 \le \exp(-\bra I \ket_0),
\label{jensen}
\end{equation}
we obtain 
\begin{equation} 
\bra I \ket_0 \ge 0.
\label{jinq}
\end{equation}
This inequality suggests that the irreversible information loss
$I$ has a certain relation with thermodynamic irreversibility.
One may find that the argument has some similarity with that 
by Jarzinski.\cite{Jarz} 

In order to discuss the convergence of $I(\tau,\Gamma(0))$ 
for  $\tau \rightarrow \infty$,  we evaluate the value of
\begin{equation}
\lim_{\tau \rightarrow \infty} \pder{I}{\tau}=
\lim_{\tau \rightarrow \infty} 
[h_a(\tau,\Gamma(0))-\tilde h_a(-\tau,\tilde \Gamma(-\tau))].
\end{equation}
Since the value of $\alpha$ is changed in a finite time interval,
$\wedge_{i=1}^{N_p}{\cal T}(t,0;\Gamma(0)) \xi_i(\Gamma(0))$ 
approaches to the unstable manifold around $\Gamma(t)$ 
when $t \gg \tau_f$.
We thus  obtain
\begin{eqnarray}
\lim_{\tau \rightarrow \infty} 
[h_a(\tau,\Gamma(0))-\tilde h_a(-\tau,\tilde \Gamma(-\tau))] 
&= &
\lim_{\tau \rightarrow \infty} 
[h(\Gamma(\tau))-h(\tilde \Gamma(-\tau))] \\ 
&=& {d \over dt}o(N), 
\end{eqnarray}
where we have used Eq. (\ref{conv0}). We further assume that this 
convergence is so fast that the time integration becomes a finite 
value. Then,  there exists a function $\bar I$ such that 
\begin{equation}
\lim_{\tau \rightarrow \infty} \lim_{N\rightarrow \infty} 
{1 \over N}I(\tau, \Gamma(0))=\bar I(\Gamma(0)).
\label{convI}
\end{equation}

\subsection{Most probable value}

We define the most probable value of $\bar I(\Gamma(0))$ based on
the assumption of the large deviation property:

Let $\Pi_I(\psi;\Sigma_0,\alpha())d\psi $ be a probability  
such that $I(\tau,\Gamma(0))/N$ 
takes a value in $[\psi, \psi+d\psi]$  when the initial equilibrium  
state  $\Sigma_0$ and the protocol of the parameter change 
$\alpha()$ are given.  Then, $\Pi_I$ is written in the form
\begin{equation}
\Pi_I(\psi;\Sigma_0,\alpha()) \sim 
\exp(-N \phi_I(\psi;\Sigma_0,\alpha())),
\label{LDpsi}
\end{equation}
in the thermodynamic limit.

The probability density $\Pi_I$ is induced from the microcanonical
measure for the initial conditions on $\Sigma_0$. 
The rate function $\phi_I$ is a convex  and non-negative function, 
and the most probable value $\bar I_*$  satisfies
\begin{equation}
\phi_I( \bar I_*;\Sigma_0,\alpha())=0.
\label{mpvp}
\end{equation}
Here,  Eq. (\ref{jinq}) leads to the inequality
\begin{equation}
\bar I_* \ge 0
\label{mvpp}
\end{equation}
for an arbitrary most probable process. 

We then try to find a relation between $\bar I_*$ and $\Delta S_{\rm B}$.
First, let us  note that  $\bra I \ket_0$ is rewritten  as
\begin{equation}
\bra I \ket_0 
= -\int_{\Sigma_0} 
\mu_{\rm mc}(\Delta_\epsilon (\Gamma(0));\Sigma_0) \log 
{\tilde \mu(\Delta_\epsilon (\tilde \Gamma(-\tau));\Upsilon_\tau)
\over 
\mu_{\rm mc}((\Delta_\epsilon (\Gamma(0));\Sigma_0)},
\label{I:rew}
\end{equation}
where we have used  Eq. (\ref{mrev}).
We consider the thermodynamic limit in the expression Eq. (\ref{I:rew}),
although the argument is not completely formalized yet.

When the most probable process $\Sigma_0 \fwdp \Sigma_1$ is 
realized,  $\Upsilon_\tau \cap \Sigma_1$ is dominant in 
$\Upsilon_\tau$ with respect to the microcanonical measure
on $\Sigma_0$. Also, from the mixing property, $\Upsilon_\tau 
\cap \Sigma_1$ may be identified with $\Sigma_1$ in a coarse
graining description of the phase space. Thus, we can expect that
$\tilde \mu( \Delta_\epsilon(\tilde \Gamma(-\tau));\Upsilon_\tau)$ 
in Eq. (\ref{I:rew}) may be  replaced by 
$\mu_{\rm mc}(\Delta_\epsilon( \tilde  \Gamma(-\tau));\Sigma_1)
\exp(o(N)),
$
in an appropriate limit of  small $\epsilon$, large $\tau$, and
large $N$. When this expectation is valid, we can rewrite 
Eq. (\ref{I:rew}) as 
\begin{eqnarray}
\bra I \ket_0 
&=& -\int_{\Sigma_0} 
\mu_{\rm mc}(\Delta_\epsilon (\Gamma(0));\Sigma_0) \log 
{\mu_{\rm mc}(\Delta_\epsilon (\tilde \Gamma(-\tau));\Sigma_1)
\over 
\mu_{\rm mc}((\Delta_\epsilon (\Gamma(0));\Sigma_0)}
+o(N) ,
\label{I:rew2} \\
&=& -\int_{\Sigma_0} 
\mu_{\rm mc}(\Delta_\epsilon (\Gamma(0));\Sigma_0) \log 
{|\Sigma_0| \over |\Sigma_1|}+o(N).
\label{I:rew3}
\end{eqnarray}
Thus, we obtain 
\begin{equation}
\bar I_*= \lim_{N \rightarrow \infty}
{1 \over N}  \log {|\Sigma_1| \over |\Sigma_0| }.
\label{I:rew4}
\end{equation}
This equality Eq. (\ref{I:rew4}) shows that the value of 
$\bar I_*$ is determined by initial and final states 
irrespective of details of the process $\Sigma_0 \fwdp \Sigma_1$.
Furthermore, from Eqs. (\ref{ondo}) in  Section 3.2, we can express
$\bar I_*$ in the form
\begin{eqnarray}
\bar I_* &=& \lim_{N\rightarrow \infty}
  {1 \over N}  \log {\Omega_1 \over \Omega_0 },\\
         &=& \lim_{N \rightarrow \infty} 
  {1 \over N} [S_{\rm B}(\Sigma_1)-S_{\rm B}(\Sigma_0)].
\label{I:rew5}
\end{eqnarray}
That is, the most probable value of the irreversible 
information loss is equal to the change of the Boltzmann entropy
per a unit degree. 
Also, the inequality Eq. (\ref{mvpp}) can be read as 
\begin{equation}
S_{\rm B}(\Sigma_1) \ge S_{\rm B}(\Sigma_0)+o(N)
\label{epri:b}
\end{equation}
for an arbitrary most probable process $\Sigma_0 \fwdp \Sigma_1$. 
This implies that the Boltzmann entropy satisfies
the entropy principle in thermodynamics.

\subsection{Fluctuation theorem}

Let us consider a probability 
$\Pi_I(\psi;\Upsilon_\tau,\tilde\alpha()) d\psi$ 
such that  the irreversible information loss $\tilde I$ 
takes a value  within $[N\psi,N(\psi+d\psi)]$
when the measure  of the initial conditions is assumed to be 
$\tilde \mu$ and the time-reversed protocol
$\tilde \alpha()$ is given.  
$\Pi_I(\psi;\Upsilon_\tau,\tilde \alpha()) d\psi$ is written  as
\begin{equation}
\tilde \Pi(\psi;\Upsilon_\tau,\tilde \alpha()) d\psi 
=
\int_{\Upsilon_\tau} 
\tilde \mu(\Delta_\epsilon(\tilde \Gamma(-\tau)); \Upsilon_\tau)
{\bf E}(\psi \le \tilde I(\tau, \tilde \Gamma(-\tau))/N \le 
\psi+d\psi), 
\label{166}
\end{equation}
where ${\bf E}(*)$ takes the value 1 when the statement $*$ 
is true.  The right-hand side of this expression is 
rewritten in the following way.
\begin{eqnarray}
&\phantom{=} &\int_{\Upsilon_\tau} 
\tilde \mu( \Delta_\epsilon(\tilde \Gamma(-\tau));\Upsilon_\tau)
{\bf E}(-\psi -d\psi \le I(\tau, \Gamma(0) ) /N \le -\psi) \\
&=&
\exp(N\psi)
\int_{\Sigma_0} 
 \mu_{\rm mc} (\Delta_\epsilon(\Gamma(0)))
{\bf E}(-\psi -d\psi \le I(\tau, \Gamma(0) )/N \le 
-\psi) \\
&=&
\exp(N\psi) \Pi_I(-\psi;\Sigma_0,\alpha())d\psi,
\end{eqnarray}
where we have used Eq. (\ref{mrev}) in order to obtain the second line.
Therefore, we obtain
\begin{equation}
{\Pi_I(\psi;\Sigma_0,\alpha()) \over  
\tilde \Pi_I(-\psi;\Upsilon_\tau,\tilde \alpha())}
=\exp(N\psi).
\label{cohen}
\end{equation}
This may be called the fluctuation theorem in Hamiltonian systems
with a time-dependent parameter.

We could not derive a useful expression of Eq. (\ref{cohen}) 
in the thermodynamic limit. We explain the reason.
Suppose that  the most probable process 
$\tilde \Sigma_1 \fwdp \Sigma_0$ is realized by the time-reversed 
protocol 
$\tilde \alpha()$. Then, the dominant region of $\Upsilon_\tau$ with 
respect to the measure $\tilde \mu$, which contributes 
$\Pi_I(\psi;\Upsilon_\tau,\tilde \alpha())$ much, is around the 
energy surface $\tilde \Sigma_1$. However, if $\Upsilon_\tau$ was 
replaced by $\tilde \Sigma_1$, $\psi$ in Eq. (\ref{cohen}) could  
not be  substituted by, for example, the most probable value 
$\bar I_*$ for $\Sigma_0 \fwdp \Sigma_1$. 
This loses the significance of Eq. (\ref{cohen}).

When we are concerned with  an infinitely small step process, 
we can derive the fluctuation-response relation from Eq. (\ref{cohen}).
In such a process,  $\Upsilon_\tau$ may be replaced by $\Sigma_0$
at the lowest order approximation.  Then, 
since $\Pi_I$ may be approximated by a Gaussian distribution
for large $N$, we can write
\begin{eqnarray}
\log \tilde \Pi_I(-\psi;\Sigma_0,\tilde \alpha())
&=& -N{(\psi+\bar I_*')^2 \over 2 \sigma'^2}+o(N), \\
\log  \Pi_I(\psi;\Sigma_0, \alpha()) 
&=& -N{(\psi-\bar I_*)^2 \over 2 \sigma^2 }+o(N).
\end{eqnarray}
The fluctuation theorem Eq. (\ref{cohen}) leads to  
\begin{equation}
-N{(\psi-\bar I_*)^2 \over 2 \sigma^2}
+N{(\psi+\bar I_*')^2 \over 2 \sigma'^2}
=N\psi+o(N).
\label{fluc:2}
\end{equation}
Since $\bar I_*$ and $\bar I_*'$ are infinitely small,
Eq. (\ref{fluc:2})  may be valid for arbitrary $\psi$
in a finite range including $\psi=0$. Thus, we have
\begin{eqnarray}
\bar I_*'&=& \bar I_*,\\
\sigma'^2&=& \sigma^2 .
\end{eqnarray}
Substituting these equalities into Eq. (\ref{fluc:2}),
we obtain 
\begin{eqnarray}
N \bar I_* &=& {N \over 2} \sigma^2 \\
         &=& {N^2 \over 2} \bra (\delta \psi)^2 \ket \\
         &=& {1 \over 2} \bra (I- N \bar I_*)^2 \ket. 
\label{fluc:3}
\end{eqnarray}
Comparing Eq. (\ref{fluc:3}) with  Eq. (\ref{FDT}) in Section 3.4,
we find that this result is consistent with Eq. (\ref{I:rew5}).

\section{Excess information loss}

At each point $\Gamma(t)$ in  a trajectory segment, 
$\{ \Gamma(t),  0 \le t \le \tau \}$, we  can consider
a  Hamiltonian  system defined on the energy surface
$\Sigma(t)$  by  {\it virtually} fixing the parameter value
to $\alpha(t)$. Then, as discussed in Section 4.4,
we can calculate the  information loss rate, $h(\Gamma;\Sigma(t))$ 
at $\Gamma \in  \Sigma(t)$  in this virtual Hamiltonian system.
We define the 
{\it excess information loss rate} as
\footnote{We obtained the idea of the excess information loss from
a paper by Oono and Paniconi,\cite{Oono}  where they defined the
excess heat in constructing steady state thermodynamics.}
\begin{equation}
h_{\rm ex}(t,\Gamma(0))=h_{\rm a}(t,\Gamma(0))
-h(\Gamma(t);\Sigma(t)).
\end{equation}
Further, the excess information loss $H_{\rm ex}$ is defined as
the time integration of $h_{\rm ex}$
\begin{equation}
H_{\rm ex}(\tau,\Gamma(0)) 
= \int_0^\tau dt h_{\rm ex}(t,\Gamma(0)).
\end{equation}
Similarly, the excess information loss rate at $\tilde \Gamma(-t)$
in the  time-reversed trajectory  is  given  by
\begin{equation}
\tilde h_{\rm ex}(-t,\tilde \Gamma(-\tau))
=\tilde h_{\rm a}(-t,\tilde \Gamma(-\tau))
-h(\tilde \Gamma(-t);\Sigma(t)),
\end{equation}
where $\tilde \Gamma(-t) \in \Sigma(t)$, and 
the excess information loss in the time-reversed trajectory
is written as 
\begin{equation}
\tilde H_{\rm ex}(\tau,\tilde \Gamma(-\tau)) 
=  \int_0^\tau dt \tilde h_{\rm ex}(-t,\tilde \Gamma(-\tau)) .
\end{equation}

Using these quantities, we rewrite the irreversible information loss $I$ as
\begin{eqnarray}
I(\tau,\Gamma(0)) 
&=& \int_0^\tau  dt
[h_{\rm a}(t,\Gamma(0))
-\tilde h_{\rm a}(-t,\tilde \Gamma(-\tau))]+o(N), \\
&=& \int_0^\tau  dt
[h_{\rm ex}(t,\Gamma(0))-
\tilde h_{\rm ex}(-t,\tilde \Gamma(-\tau))]+o(N), \\
&=& H_{\rm ex}(\tau,\Gamma(0))-
\tilde H_{\rm ex}(\tau,\tilde \Gamma(-\tau))+o(N),
\label{i2hex}
\end{eqnarray}
where we have used the equality 
\begin{equation}
h( \Gamma(t);\Sigma(t))=
h(\tilde \Gamma(-t);\Sigma(t))+{d \over dt}o(N).
\end{equation}
(See Eq. (\ref{conv0}).)  Further, through the definition of 
$H_{\rm ex:rev}$
\begin{equation}
H_{\rm ex:rev}(\tau,\Gamma(0)) 
= {1\over 2} [H_{\rm ex}(\tau,\Gamma(0))
+\tilde H_{\rm ex}(\tau,\tilde \Gamma(-\tau))],
\end{equation}
Eq. (\ref{i2hex}) becomes 
\begin{equation}
{1 \over 2} I(\tau,\Gamma(0))=
H_{\rm ex}(\tau,\Gamma(0))-H_{\rm ex:rev}(\tau,\Gamma(0))+o(N). 
\label{i2hex2}
\end{equation}

In the viewpoint of numerical calculation, the excess information loss 
is a more tractable quantity than the irreversible information loss,
because $H_{\rm ex}(\tau, \Gamma(0))$ converges to a certain value
$H_{\rm ex}(\infty, \Gamma(0))$ when $\tau \rightarrow \infty$.
This convergence can be expected  from a fact that 
$h_{\rm ex}(t,\Gamma(0))$ converges to $0$ when 
$t-\tau_f \rightarrow \infty$. 
Similarly, we expect that $\tilde H_{ex}(\tau, \tilde \Gamma(-\tau))$ 
converges to a certain value 
$\tilde H_{\rm ex}(\infty, \tilde \Gamma(-\infty))$
when $\tau_i \rightarrow \infty$. (Note that $\tau \rightarrow \infty$
when $\tau_i \rightarrow \infty$.) Also, 
$H_{\rm ex:rev}(\infty,\Gamma(0))$ is determined.


Let us consider the  average over the initial conditions
sampled from the microcanonical ensemble on the energy surface
$\Sigma_0$. From Eq. (\ref{i2hex2}), we have
\begin{equation}
{1 \over 2} N \bar I_*=
\bra H_{\rm ex} \ket_0 - \bra H_{\rm ex:rev} \ket_0+o(N)
\label{s2hex2}
\end{equation}
in the thermodynamic limit, where $\bra H_{\rm ex} \ket_0$ is the
average of $H_{\rm ex}(\infty,\Gamma(0))$. Since 
$H_{\rm ex}(\infty,\Gamma(0))$ can be 
obtained numerically without referring the time-reversed trajectory,
$\bra H_{\rm ex} \ket_0$ is a directly measurable quantity.
Although $\bra H_{\rm ex:rev} \ket_0$ is not easily obtained
numerically,  this may be expected to have a certain
relation to the {\it quasi-static excess information loss}
$H_{\rm ex:qs}$, which  is  defined as
\begin{equation}
H_{\rm ex:qs}=\int_0^\infty dt \der{\alpha}{t}  \Phi(\Sigma(t)),
\label{hexqs:def}
\end{equation} 
where the quantity $\Phi(\Sigma(t)) d\alpha$ is the excess information 
loss calculated  under the assumption that the equilibrium state 
is {\it virtually} realized  at each time $t$ along the trajectory 
$\Gamma()$. (Recall a similar discussion
below Eq. (\ref{final}).) Notice that $H_{\rm ex:qs}$ becomes the real 
excess information loss when the process is quasi-static. 

We are going to discuss the relation between $H_{\rm ex:rev}$ and 
$H_{\rm ex:qs}$. We consider a step process realized by an infinitely 
small parameter 
change $\alpha \rightarrow \alpha+\Delta \alpha$ at $t=\tau_i$.
Then, the quantity $\Phi(\Sigma_0)$ is given by
\begin{equation}
\bra H_{\rm ex} \ket_0= \Phi(\Sigma_0) \Delta \alpha
+O( (\Delta \alpha)^2) .
\end{equation}
Also, using Eqs. (\ref{ent:step:N}) and (\ref{I:rew5}), we obtain
\begin{equation}
\bar I_* =O( (\Delta \alpha)^2).
\end{equation}
Therefore, we find 
\begin{equation}
\bra  H_{\rm ex:rev} \ket_0= \Phi(\Sigma_0) \Delta \alpha
+O( (\Delta \alpha)^2).
\label{rev1st}
\end{equation}
Since a quasi-static process can be realized by repeating
an infinite number of infinitely small step processes,
$\bra  H_{\rm ex:rev} \ket_0$ for a quasi-static process 
$\Sigma_0 \qsp \Sigma_1$ is written as
\begin{equation}
\bra  H_{\rm ex:rev} \ket_0
= \int_{\alpha_0}^{\alpha_1}  d\alpha  \Phi(\Sigma_{(\alpha)}), 
\label{revqs0}
\end{equation}
where $\Sigma_{(\alpha)}$ is the equilibrium state such that 
\begin{equation}
\Sigma_0=\Sigma_{(\alpha_0)} \qsp  \Sigma_{(\alpha)}.
\end{equation}
Then, Eq. (\ref{revqs0}) implies
\begin{equation}
\bra H_{\rm ex:rev} \ket_0 = \bra H_{\rm ex:qs} \ket_0.
\label{revqs}
\end{equation}
Note however  that  the validity of Eq. (\ref{revqs}) is ensured 
only for  quasi-static processes.  Nevertheless, we assume that 
Eq. (\ref{revqs}) holds at least near quasi-static processes.   Based on the
assumption, the right-hand side of Eq. (\ref{s2hex2}) is calculated
numerically without referring time-reversed trajectories, and
from Eqs. (\ref{I:rew5}), (\ref{s2hex2}) and (\ref{revqs}), we 
obtain the expression
\begin{equation}
{1 \over 2}\bra \Delta S_{\rm B} \ket_0=
\bra H_{\rm ex} \ket_0 - \bra H_{\rm ex:qs} \ket_0+o(N).
\label{7:final}
\end{equation}

\subsection{Minimum principle}

Applying the inequality Eq. (\ref{mvpp}) to the expression 
Eq. (\ref{s2hex2}), we obtain
\begin{equation}
\bra H_{\rm ex} \ket_0 \ge  \bra H_{\rm ex:rev} \ket_0 +o(N).
\label{mp}
\end{equation}
This inequality implies that {\it the  excess information loss must not
be lower than its reversible part.} Such a phrase reminds us the 
minimum work principle in  thermodynamics with an isothermal 
environment, which states that the work done by external agents 
must not be lower than the quasi-static work. As discussed in the 
previous subsection, $H_{\rm ex:rev}$ may be related to the 
quasi-static excess information loss. Therefore, the analogy 
with the minimum work principle may be expected more and 
Eq. (\ref{mp}) may be regarded as the minimum excess information loss 
principle.  However, we do not yet understand the 
significance of the  inequality Eq. (\ref{mp}).  We expect
that the analysis of subsystems may provide us a further insight 
for Eq. (\ref{mp}). This will be a future problem.

\subsection{Expression of $\Phi$}

In this subsection, we derive an expression of $\Phi$ 
in terms of  Lyapunov vectors. Suppose that the value of  $\alpha$ 
is changed instantaneously from $\alpha_0$ to $\alpha_0+\Delta \alpha$
at  time $t=0$.  The  trajectory $\Gamma( )$ is not differentiable 
at $t=0$. We consider the excess information loss for $\Gamma()$ 
\begin{equation}
H_{\rm ex}(\infty,\Gamma(0))=
\int_0^{\infty} dt [h_{\rm a}(t;\Gamma(0))- h(\Gamma(t);\Sigma_1)],
\label{hex0}
\end{equation}
where $\Sigma_1$ is an energy surface after the parameter change.

Let  $\{ {\xi}_i^{(0)},\  1 \le i \le 2N \}$ be a set of 
Lyapunov vectors at $\Gamma(0)$ in the energy surface $\Sigma_0$.
We then define a set of vectors $\{ {a}_i(t),\  1 \le i \le 2N \}$ 
in the tangent space at $\Gamma(t)$  as
\begin{equation}
 {a}_i(t)={\cal T}(t,0;\Gamma(0)) \xi_i^{(0)},
\label{defa}
\end{equation}
where ${\cal T}(t,0;\Gamma(0))$ is  the linearized evolution
map along the trajectory in the energy surface $\Sigma_1$.
Note that $ {a}_i$ is not the Lyapunov vector  at $\Gamma(t)$,
because $ {\xi}_i^{(0)}$ is {\it not} the 
Lyapunov vector in the energy surface $\Sigma_1$.

The $i$-th Lyapunov vector along the trajectory 
$\{ \Gamma(t), \ 0 \le t \le \infty\}$ in the energy 
surface $\Sigma_1$ is  denoted by $ {\xi}_i^{(1)}(\Gamma(t))$.  
The expansion factor $\Lambda_i(t,\Gamma(0))$ satisfies
\begin{equation}
{\cal T}(t,0;\Gamma(0)) {\xi}_i^{(1)}(\Gamma(0))=\Lambda_i(t,\Gamma(0))
 {\xi}_i^{(1)}(\Gamma(t)).
\end{equation}
Using $a_i$ and $\Lambda_i$, we can write Eq. (\ref{hex0}) as
\begin{equation}
H_{\rm ex}(\infty,\Gamma(0))=
\lim_{t \rightarrow \infty}
[\log |\wedge_{i=1}^{N_p}  {a}_i(t)|-
\sum_{i=1}^{N_p}\log \Lambda_i(t,\Gamma(0)) ].
\label{hex1}
\end{equation}

Let us evaluate the right-hand side of Eq. (\ref{hex1}).
We expand $ {\xi}_i^{(0)}$ by the set of Lyapunov vectors
$\{  {\xi}_i^{(1)}(\Gamma(0)), \ 1 \le i \le 2N \} $
in such a way that 
\begin{equation}
 {\xi}_i^{(0)}=\sum_{j=1}^{2N} Q_{ij}  {\xi}_j^{(1)}(\Gamma(0)).
\label{defQ}
\end{equation}
Here, the matrix ${\cal Q}$ is defined at $\Gamma(0)$ and depends
on $\Delta \alpha$. 
Then, from Eqs. (\ref{defa}) and (\ref{defQ}),
$ {a}_i(t)$ is expanded in the form
\begin{equation}
 {a}_i(t)=\sum_{j=1}^{2N} 
Q_{ij}  \Lambda_j(t,\Gamma(0)) {\xi}_j^{(1)}(\Gamma(t)).
\end{equation}
Using this expression, we write $|\wedge_{i=1}^{N_p}  {a}_i(t)| $ 
as 
\begin{equation}
\left\vert \sum_{(j_1,\cdots,j_{N_p})}
\left[ \prod_{k=1}^{N_p} Q_{kj_k} \right] 
\left[ \prod_{k=1}^{N_p} \Lambda_{j_k}(t,\Gamma(0))\right] 
\left[\wedge_{k=1}^{N_p}  {\xi}_{j_k}^{(1)}(\Gamma(t))\right] 
\right\vert,
\end{equation}
where the index $j_k$ varies from $1$ to $2N$.
When $t$ is sufficiently large, the contribution from
the unstable directions becomes dominant. Thus, for  sufficiently
large $t$, we derive
\begin{eqnarray}
\log | \wedge_{i=1}^{N_p}  {a}_i(t)| 
&\simeq &
\log 
\left\vert
\sum_{j_1=1}^{N_p}\cdots \sum_{j_{N_p}=1}^{N_p}
{\rm sgn}(j_1,\cdots,j_{N_p}) \left[\prod_{k=1}^{N_p} Q_{kj_k}\right] 
\right. \\
&\cdot & \left.\left[\prod_{k=1}^{N_p} \Lambda_{k}(t,\Gamma(0))\right] 
\left\vert \wedge_{i=1}^{N_p} {\xi}_{i}^{(1)}(\Gamma(t)) \right\vert 
\right\vert \\
&\simeq &
\log {\rm det}_+ {\cal Q}
+ \sum_{i=1}^{N_p} \log\Lambda_{i}(t,\Gamma(0))] ,
\label{191}
\end{eqnarray}
where 
\begin{equation}
{\rm det}_+ {\cal Q}=
\sum_{j_1=1}^{N_p}\cdots \sum_{j_{N_p}=1}^{N_p}
{\rm sgn}(j_1,\cdots,j_{N_p}) Q_{1j_1}\cdots Q_{N_p,j_{N_p}}
\end{equation}
and  we have used the normalization condition
$ |\wedge_{i=1}^{N_p}  {\xi}_{i}^{(1)}(\Gamma(t)) |=1$.
Finally, substituting Eq. (\ref{191})  to Eq. (\ref{hex1}), we obtain
\begin{equation}
H_{\rm ex}(\infty,\Gamma(0))=\log {\rm det}_+ {\cal Q}.
\end{equation}
Therefore, from the definition of $\Phi(\Sigma_0)$,
we have
\begin{equation}
\Phi(\Sigma_0)=\lim_{\Delta \alpha \rightarrow 0} 
{ \bra \log {\rm det}_+ {\cal Q} \ket_0 \over \Delta \alpha}.
\end{equation}
This expression shows that $\Phi$ takes a non-zero value
when the unstable manifold varies linearly for the infinitely
small parameter change.

\section{Numerical experiments}

Until now, we have developed the theoretical arguments. 
However, one may point out that these lack 
the mathematical rigorousness. We then present  evidences  by numerical 
experiments so as to  confirm the validity of the theoretical arguments.

As a direct experimental test of our theory, we should check 
the relation 
between the irreversible information loss  and  the Boltzmann entropy. 
However, unfortunately, we do not yet complete this test,
because it is hard to measure numerically the irreversible information loss.
The reasons of this hardness are as follows.

First, the time-reversed trajectory is needed in calculation of 
$I(\tau,\Gamma(0))$.
This fact causes a delicate problem: Suppose that we obtain 
numerically a trajectory segment $\{ \Gamma(t), \ 0 \le t \le \tau\}$. 
Since the system is chaotic, this trajectory is not an approximation 
of the true trajectory starting from the initial condition $\Gamma(0)$.
However, when a pseudo-orbit tracing property is valid in the system, 
there is a true trajectory which is closed to the trajectory segment 
obtained numerically.  Then, when we integrate the time-reversed 
equations of motion with the initial condition $\tilde \Gamma(-\tau)$, 
the trajectory obtained numerically deviates from 
$\{ \tilde \Gamma(t), \ -\tau \le t \le 0\}$ due to the orbital 
instability. Therefore, in order to obtain the time-reversed trajectory 
$\{ \tilde \Gamma(t), \ -\tau \le t \le 0\}$, we must store the 
data of the original trajectory.

Second, even if we obtain numerically $I(\tau,\Gamma(0))$,
 it does not converge to a value for $\tau 
\rightarrow \infty$,  because only the extensive part of $I(\tau,\Gamma(0))$ 
converges. Therefore, it is not easy to choose the 
value of $\tau$ in numerical calculation. In principle, we have only to 
choose large $N$ so that the arbitrariness becomes less. 
However, we need much more time to study the systems with larger $N$.

In this paper,  instead of the irreversible information loss, we 
discuss numerically the excess information loss, with particularly 
focusing on  Eq. (\ref{7:final}). Since $H_{\rm ex}(\tau,\Gamma(0))$ 
converges to $H_{\rm ex}(\infty,\Gamma(0))$ when $\tau \rightarrow \infty$,
the numerical calculation  of the excess information loss may be simpler 
than that of the irreversible information loss. Also, the time-reversed 
trajectory is not needed in the calculation of $H_{\rm ex}$.

\subsection{Model}

A system consisting of many molecules with  
short-range repulsive interaction may be the most realistic model 
to study thermodynamic irreversibility. 
However, since we are concerned with a universal aspect of 
irreversibility, the choice of the system does not matter.  
Simpler models may be better to us. 
That is the reason why we study numerically the Fermi-Pasta-Ulam (FPU) 
model.\cite{FPU}  The Hamiltonian of the FPU model is given by
\begin{equation}
H(\{q _i\}, \{p_i \};g)= 
\sum_{i=1}^N [{1 \over 2} p_i^2+
{1\over 2}(q_{i+1}-q_{i})^2+{g \over 4}(q_{i+1}-q_{i})^4],
\label{hamil}
\end{equation}
where the value of $g$ is changed in time.  That is, $\alpha()$ 
in previous sections is identified to $g()$ in this section.

The evolution equations for  $(\{q_i\},\{p_i\})$ are written as
\begin{eqnarray}
{d q_i \over dt}  &=& p_i, 
\label{evol:q} \\
{d p_i \over dt}  &=& (q_{i+1}-q_{i})+g(q_{i+1}-q_{i})^3
-(q_{i}-q_{i-1})-g(q_{i}-q_{i-1})^3.
\label{evol:v}
\end{eqnarray}
We assume periodic boundary conditions, 
$q_0 = q_{N}$ and $q_{N+1}= q_{1}$ in Eq. (\ref{evol:v}).
Under the boundary conditions, $\sum_i p_i$ is a conserved
quantity. We assume that the value of $\sum_i p_i$ is zero,
for simplicity. Then,  $\sum_i q_i$  also becomes a
conserved quantity. We  assume that the value of
$\sum_i q_i$ is zero.  In the remaining part of this section,
the energy surface with the condition $\sum_i p_i=\sum_i q_i=0$
is simply called the energy surface. 
We solve numerically
Eqs. (\ref{evol:q}) and (\ref{evol:v}) by the 4-th order symplectic 
integrator method\cite{symp} with a time step $\delta t=0.005$.
Since  we are concerned with the thermodynamic limit, we  check 
the $N$ dependence of our conclusions.

\subsection{Lyapunov analysis}

In this subsection, we assume that $g$ takes a constant value, say
$g_0$. Let $E_0$ be the energy. When $ E_0g_0$ is sufficiently large, 
the system with large $N$ exhibits high-dimensional chaos.
As an example of such parameter value set, $(E_0,g_0)=(1.0, 10.0)$ 
is assumed. 

\begin{figure}
\epsfxsize=14cm
\centerline{\epsfbox{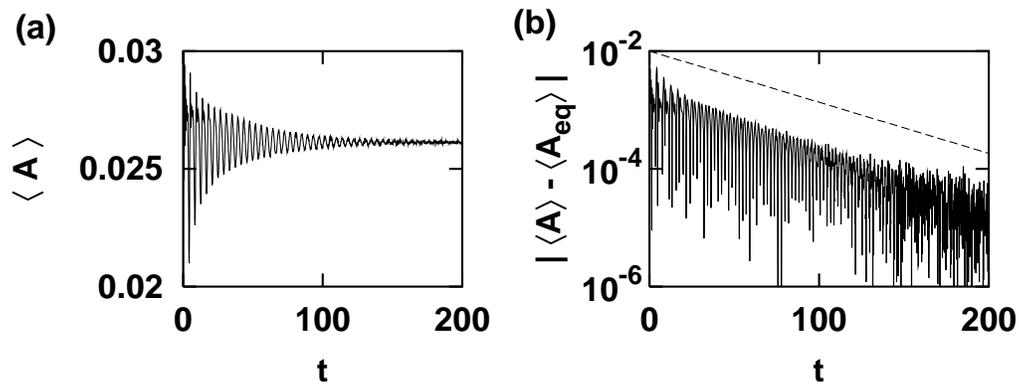}}
\caption{Relaxation behavior of the ensemble average of $A$. $N=20$. 
The ensemble of initial conditions is made artificially with fixing 
$E$ and $A$ as $E=1$ and $A(0)=0.01$.
(b)$ \log |\bra A \ket- A_{\rm eq}|$ versus $t$. $A_{\rm eq}$ is the
equilibrium value determined by the graph (a). The dotted line
shows $|\bra A \ket- A_{\rm eq}| =\exp(-t/50)/100$.
}
\label{fig1}
\end{figure}

We first check the mixing property with respect to the 
micro-canonical measure by discussing a relaxation behavior.
(See the last paragraph of Section 2.3.)  Figure \ref{fig1} 
shows an example how the average of $A$ relaxes to the equilibrium 
value when the initial conditions are sampled from an artificial
ensemble we assumed.  
As far as we checked, we observed a similar relaxation behavior 
to the same equilibrium value for different sets of the initial conditions.
We thus conclude numerically
that the system possesses the mixing property.
Therefore, 
the ensemble of the initial conditions at $t=0$  is regarded as
the microcanonical ensemble with the energy $E_0$ when the ensemble 
is made by sufficiently long time evolution of phase space points 
sampled from a distribution absolutely continuous to the Lebesgue 
measure on the energy surface.
Here, we remark that the relaxation curve includes an oscillatory
component, while the envelop curve exhibits an exponential 
decreasing behavior. Both the period of the oscillation and 
the relaxation time seem to be larger for larger $N$.

In order to demonstrate the chaotic nature quantitatively, 
we show the Lyapunov exponents in Fig. {\ref{fig3}. Notice that 
there are two additional zero Lyapunov exponents because of the 
momentum conservation.  That is, $N_{\rm p}=N-2$. 
The convergence of orthonormal frames 
is confirmed in the way described in Section 4.2. 
Figure \ref{fig4} shows  the time evolution of the average of 
the distance 
$d({\cal F}(t,\Gamma(0)) {e}_i, {\cal F}(t,\Gamma(0)) {e}'_i)$
over initial conditions, where $i=1,2, N-3$ and $N-2$. 
We see that the distance decreases to a computational noise level
after $t=3000$. 

\begin{figure}
\epsfxsize=7cm
\centerline{\epsfbox{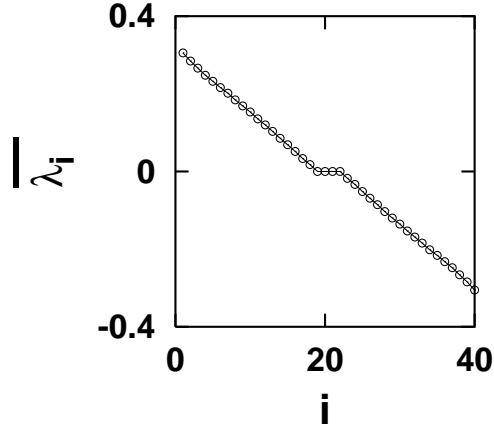}}
\caption{Lyapunov spectrum. $N=20$. 
}
\label{fig3}
\end{figure}

\begin{figure}
\epsfxsize=7cm
\centerline{\epsfbox{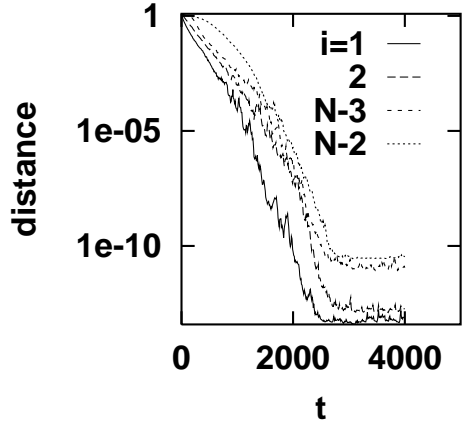}}
\caption{Convergence of a set of orthogonal unit vectors. $N=20$.
}
\label{fig4}
\end{figure}

\subsection{Boltzmann entropy}

The Boltzmann entropy $S_{\rm B}(E,g)$ is calculated numerically 
in the following way: First, according to the adiabatic theorem,
which was discussed in Section 3.1, the phase space volume enclosed by
an energy surface is conserved along quasi-static processes. Therefore,
the equality 
\begin{equation}
\Omega(E,g)= \Omega(E_*, 0)
\end{equation}
holds for the quasi-static process $(E,g) \qsp  (E_*,0)$.
Since the FPU model with $g=0$ is reduced to the harmonic oscillator 
model, $\Omega(E_*, 0)$ is given by the volume of the $2N-2$ 
dimensional sphere, and is calculated as
\begin{equation}
\Omega(E_*, 0)= c E_*^{N-1},
\label{homega}
\end{equation}
where $c$ does not depend on $E_*$.  
Thus, the Boltzmann entropy at $(E,g)$ can be evaluated as\footnote{
In our previous paper\cite{SK}, the factor in the right-hand side 
was written as $N-2$, not $N-1$. This was a mistake.}
\begin{equation}
S_{\rm B}(E,g)=S_{\rm B}(E_*,0)=(N-1)\log E_*,
\label{ent-bolt}
\end{equation}
where  an additive constant with respect to $E_*$ is omitted.

In numerical experiments, a quasi-static process 
$(E_0,g_0) \qsp (E_1,g_0+\Delta g)$ is realized by the large $\tau$
limit of the protocol 
$g(t)=g_0+\Delta g t/\tau$ for $0 \le t \le \tau$.
In Fig. \ref{fig5}, the average and deviation of $E_1$ are
plotted against $\tau$.  We find that the deviation
becomes smaller for larger $\tau$, and we may assume that 
quasi-static processes are realized  when  $\tau > 100$.
The equi-entropy curve  through $\Sigma_0=(E_0,g_0)$ in Fig. \ref{fig6}
was obtained in this way. We express the curve by
$E=E_{\rm qs}(g;\Sigma_0)$. Similarly, as shown in Fig. \ref{fig6}, 
we can draw equi-entropy curves in the $(E,g)$ space.

\begin{figure}
\epsfxsize=14cm
\centerline{\epsfbox{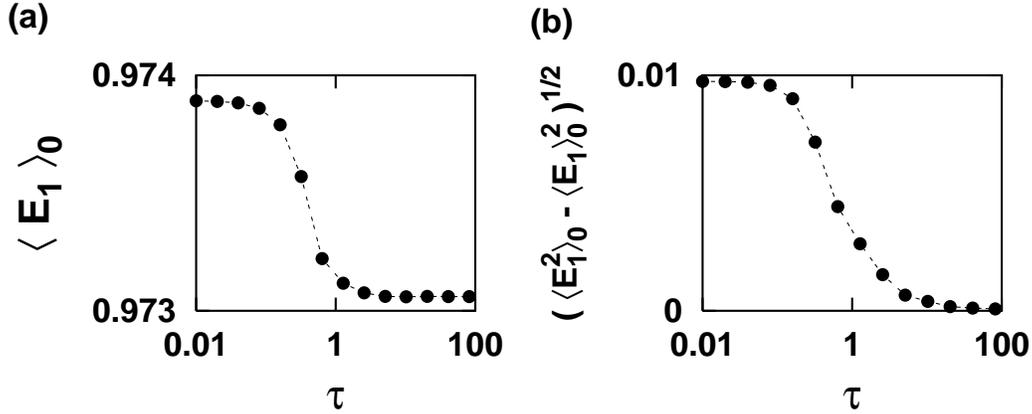}}
\caption{(a) Average value of final energy versus $\tau$. 
(b) Deviation of values of final energy versus $\tau$.
$N=20$ and $\Delta g=-1.0$.}
\label{fig5}
\end{figure}

\begin{figure}
\epsfxsize=7cm
\centerline{\epsfbox{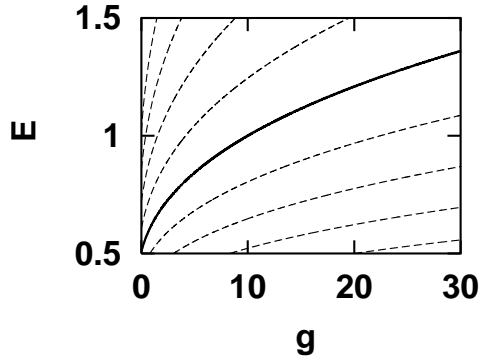}}
\caption{Equi-entropy curve through $(E_0,g_0)$ (solid curve) and 
other equi-entropy curves (dotted line). 
}
\label{fig6}
\end{figure}

Now, we consider a step process realized  by the instantaneous change
of the value of $g$, from $g_0$ to $g_1=g_0+\Delta g$, at $t=0$.
Then, the energy after the switching becomes $E_1$, 
whose value depends on the choice of the initial condition. 
The entropy difference $\Delta S_{\rm B}$ is calculated by
\begin{equation}
\Delta S_{\rm B} =S_{\rm B}(E_1,g_1)-S_{\rm B}(E_0,g_0),
\end{equation}
with the formula Eq. (\ref{ent-bolt}). In Fig. \ref{fig7}, 
the average of the entropy difference over the initial conditions, 
$ \bra \Delta S_{\rm B} \ket_0 $, is plotted  against $\Delta g$. 
This graph shows that $\bra \Delta S_{\rm B} \ket_0$ is positive.
Also, as shown in Fig. \ref{fig8}, 
the relative fluctuation of $\Delta S_{\rm B}$ becomes less
as $N$ is increased.  This implies the existence of 
the large deviation property of $\Delta S_{\rm B}$.

\begin{figure}
\epsfxsize=14cm
\centerline{\epsfbox{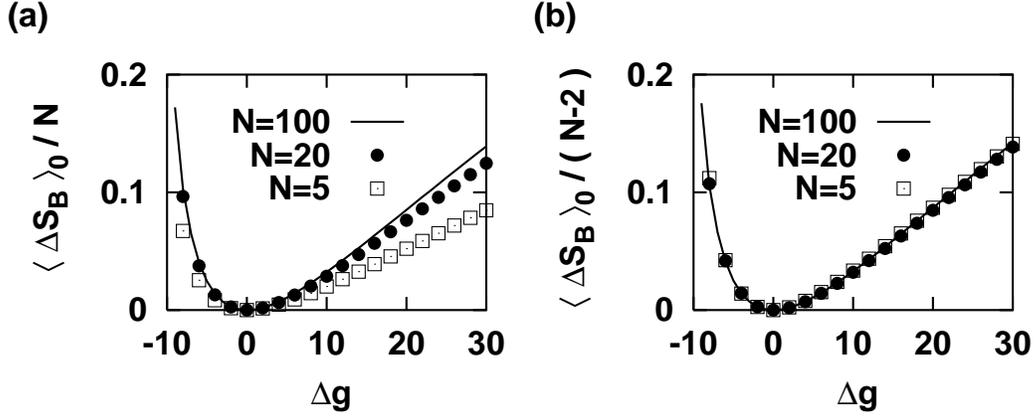}}
\caption{Entropy difference versus $\Delta g$.  
The different normalizations for the $N$ dependence are 
used in (a) and (b).
}
\label{fig7}
\end{figure}

\begin{figure}
\epsfxsize=7cm
\centerline{\epsfbox{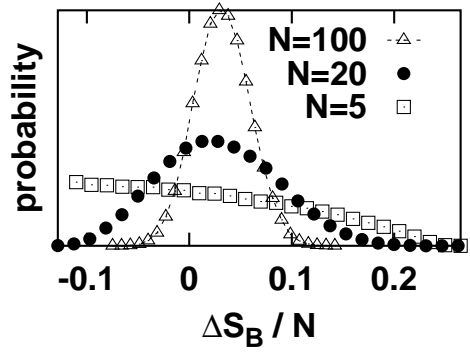}}
\caption{The probability of the entropy difference. $\Delta g=10.$
}
\label{fig8}
\end{figure}

\subsection{Excess information loss}

We show the result of numerical calculation of the excess  information 
loss for the step processes $(E_0,g_0) \rightarrow (E_1,g_0+\Delta g)$.
In Fig. \ref{fig9},   $H_{\rm ex}(t,\Gamma(0))$ for four choices of
the initial condition $\Gamma(0)$ are plotted  against $t$.
One can see that $H_{\rm ex}(\infty,\Gamma(0))$ is clearly defined. 
In Fig. \ref{fig10}, 
we show the average of $H_{\rm ex}(t,\Gamma(0))$ over the initial 
conditions chosen from the microcanonical ensemble on the energy 
surface $\Sigma_0$.  $\bra H_{\rm ex} \ket_0$ is given
as the value at $t=\infty$ in this graph.

\begin{figure}
\epsfxsize=7cm
\centerline{\epsfbox{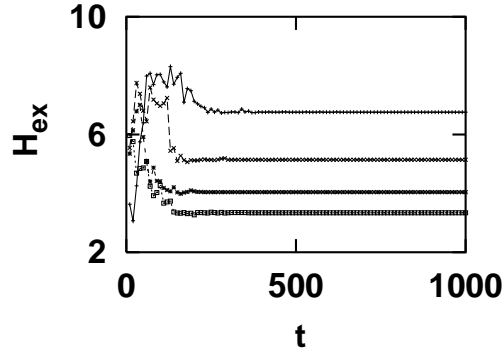}}
\caption{$H_{\rm ex}(t,\Gamma(0))$ versus $t$ for four different
initial conditions. $N=20$ and $\Delta g=20$.
}
\label{fig9}
\end{figure}

\begin{figure}
\epsfxsize=7cm
\centerline{\epsfbox{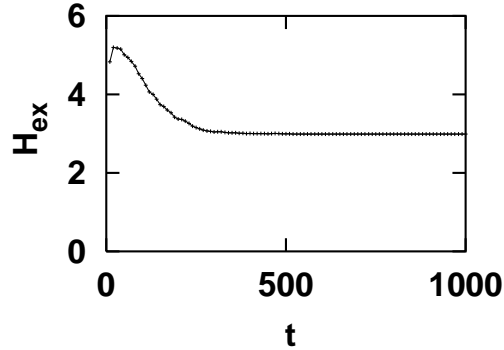}}
\caption{Average value of $H_{\rm ex}(t,\Gamma(0))$ versus $t$.
$N=20$ and $\Delta g=20$.
}
\label{fig10}
\end{figure}

In Fig. \ref{fig11},  $\bra H_{\rm ex} \ket_0 $ is  plotted against 
$\Delta g$. From this graph, we can evaluate the value of 
$\Phi(\Sigma_0)$ by the equality 
\begin{equation}
\bra H_{\rm ex} \ket_0 =\Phi(\Sigma_0) \Delta g+o(\Delta g)
\end{equation}
for $\Delta g \rightarrow 0$.  In a similar way, in principle, 
we can calculate the value of $\Phi(\Sigma)$ at each energy surface.
In particular, $H_{\rm ex:qs}$ at the step process is given by
\begin{equation}
H_{\rm ex:qs} ={1\over 2} (\Phi(\Sigma_0)+\Phi(\Sigma_1))\Delta g,
\label{hexqse}
\end{equation} 
where $\Sigma_1=(E_1,g_1)$ and  recall Eq. (\ref{hexqs:def}) for 
the definition of $H_{\rm ex:qs}$.

\begin{figure}
\epsfxsize=14cm
\centerline{\epsfbox{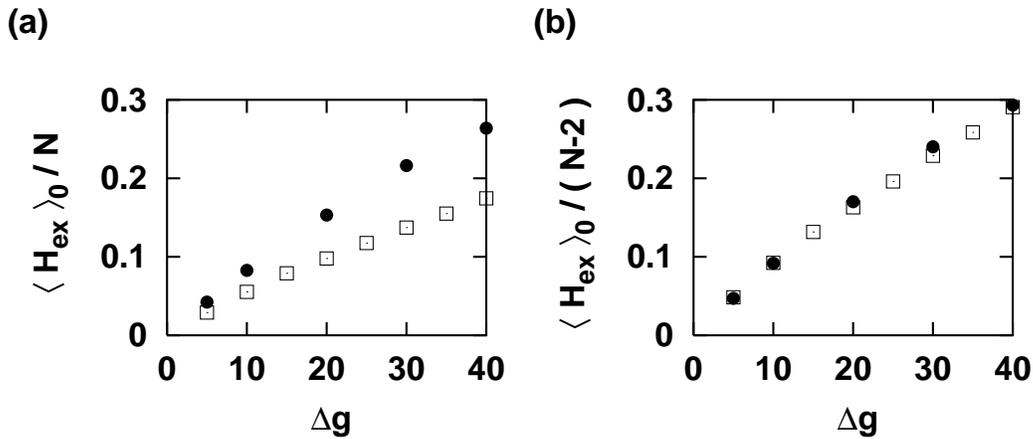}}
\caption{$\bra H_{\rm ex} \ket_0$ versus $\Delta g$.
Square and filled circle symbols represent the data for $N=5$ and $N=20$, 
respectively. The different normalizations for the $N$ dependence are 
used in (a) and (b). 
}
\label{fig11}
\end{figure}

\subsection{Main experiment}

In order to check the validity of Eq. (\ref{7:final}), we 
need to calculate  $ \bra H_{\rm ex:qs} \ket_0 $, the average of 
$H_{\rm ex:qs}$ over the initial conditions. This is written  as
\begin{equation}
\bra H_{\rm ex:qs} \ket_0 =
{1\over 2} (\Phi(\Sigma_0)+ \bra \Phi(\Sigma_1) \ket_0)\Delta g.
\end{equation} 
However, this average cannot be calculated efficiently,
because $\Sigma_1$ depends on the choice of the initial condition,
and $\Phi(\Sigma_1)$ is evaluated from a graph indicating 
$\Delta g$ versus the average of $ H_{\rm ex} $ over initial 
conditions sampled from the microcanonical ensemble on  $\Sigma_1$.  
We thus find out a way to avoid the calculation of $H_{\rm ex:qs}$.
We perform  the reversed experiments in which the parameter $g$ is 
changed from $g_1$ to $g_0$ with the initial state 
$\Sigma_0'=(E_{\rm qs}(g_1;\Sigma_0),g_1)$. Suppose that 
the process $\Sigma_0' \rightarrow \Sigma_1'$ is realized,
where $\Sigma_1'$ depends on the choice of the initial condition
on the energy surface $\Sigma_0'$.  Then, we can calculate 
the Boltzmann entropy change $\Delta S_{\rm B}$ and the excess information 
loss, $H_{\rm ex}'$,  for the process $\Sigma_0'\rightarrow \Sigma_1'$. 
These satisfy the relation
\begin{equation}
{1\over 2} \bra \Delta S_{\rm B} \ket_0' = 
\bra H_{\rm ex}'\ket_0'- \bra H_{\rm ex:qs}'\ket_0'+o(N),
\label{hex-ds:rev}
\end{equation}
where $\bra \ket_0'$ denotes the average over the initial conditions
sampled from the microcanonical ensemble on the energy surface 
$\Sigma_0'$, and $ H_{\rm ex:qs}'$ denotes the quasi-static excess 
information loss for the step process $\Sigma_0'\rightarrow \Sigma_1'$.
Here, we can prove the estimation
\begin{equation}
\bra H_{\rm ex:qs} \ket_0 + \bra H_{\rm ex:qs}' \ket_0'
=o( (\Delta g)^2).
\label{minus}
\end{equation} 

(proof) 

{}From the definition of $H_{\rm ex:qs}$, we have
\begin{equation}
\bra H_{\rm ex:qs}' \ket_0' =-
{1\over 2} (\Phi(\Sigma_0')+ \bra \Phi(\Sigma_1') \ket_0')\Delta g.
\end{equation} 
Using this equation and  Eq. (\ref{hexqse}), we obtain
\begin{eqnarray}
\bra H_{\rm ex:qs}\ket_0 +\bra H_{\rm ex:qs}' \ket_0'
&=& {1\over 2}( \Phi(\Sigma_0)+ \bra \Phi(\Sigma_1) \ket_0 ) \Delta g.\\
&-& {1\over 2} (\Phi(\Sigma_0')+ \bra \Phi(\Sigma_1') \ket_0')\Delta g.
\label{108}
\end{eqnarray}
Recalling the energy change for step processes, we expect
\begin{eqnarray}
\Sigma_0'-\Sigma_1 &\sim&  O( (\Delta g)^2), \\
\Sigma_0-\Sigma_1' &\sim&  O( (\Delta g)^2).
\end{eqnarray}
Substitution these estimation to Eq. (\ref{108}) leads to 
Eq. (\ref{minus}).

\hfill (q.e.d.)

Finally,  from Eqs. (\ref{7:final}), (\ref{hex-ds:rev}) and (\ref{minus}), 
we obtain the equality 
\begin{equation}
{1\over 2} ( \bra \Delta S_{\rm B} \ket_0 + \bra \Delta S_{\rm B} \ket_0')
=\bra  H_{\rm ex} \ket_0+\bra H_{\rm ex} \ket_0'+o(N,(\Delta g)^2).
\label{hex-ds:sum}
\end{equation}
This relation can be checked numerically. 
As shown in Fig. \ref{fig12},  Eq. (\ref{hex-ds:sum}) seems 
valid, and therefore,  our theoretical arguments turn out
to be consistent.

\begin{figure}
\epsfxsize=7cm
\centerline{\epsfbox{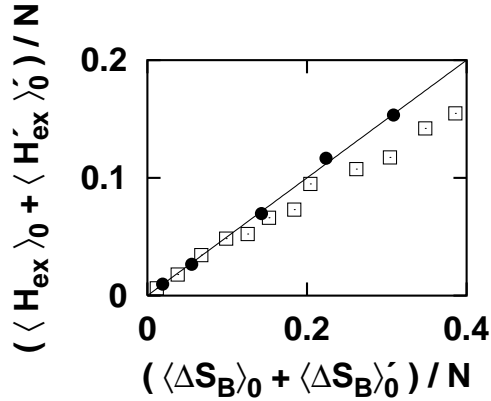}}
\caption{$\bra H_{\rm ex} \ket_0+\bra H_{\rm ex}' \ket_0'$
versus $ \bra \Delta S_{\rm B} \ket_0 + \bra \Delta  S_{\rm B} \ket_0' $.
Square and filled circle symbols represent the data for 
$N=5$ and $N=20$, respectively. The sold line shows  
$\langle H_{\rm ex} \rangle_0+\langle H_{\rm ex}' \rangle_0'=1/2[
\langle \Delta S_{\rm B} \rangle_0 + \langle \Delta  S_{\rm B} \rangle_0']. $
}
\label{fig12}
\end{figure}

\section{concluding remark}

The essence of thermodynamic irreversibility is described by the entropy 
principle. Therefore, when one discusses thermodynamic 
irreversibility from dynamical systems, the purpose is to find a state 
variable satisfying the entropy principle. 
Our arguments stand on this natural viewpoint. In this paper, 
we have found that the irreversible information loss 
leads to the state variable which satisfies the entropy  principle. 

We expect that our theory may be extended so as to apply to other 
dynamical systems without Hamiltonian.  For example,
in dissipative systems driven by external forces, a steady state is 
realized. The fluctuation properties have been discussed extensively.
However, an attempt of the construction of  state theory seems scarcely. 
Although Oono and Paniconi have proposed an operational way to construct 
a state variable in  steady state thermodynamics, its validity is not 
confirmed.\cite{Oono} We will attempt to construct non-equilibrium
thermodynamic functions from dynamical system models by studying
fluctuation properties.

Much variety of nonlinear dynamics have been known in the context 
of fluid systems, granular systems, chemical reaction systems, 
and biological systems. In these systems, the notion of  entropy 
is not self-evident, and hardly connected to that in thermodynamics.
Even in such systems, it may be important to characterize a state 
of the systems in terms of state functions representing a relation 
between states. In particular, one of the most serious questions 
in science  may be the boundary of the living state. 
One may ask why we cannot restore the living state when a life 
is dead. This is not the problem of  thermodynamic irreversibility, 
but there is a sort of irreversibility in a biological world. However, 
the question is too general to be discussed scientifically. 
We should argue more restricted phenomena  related to this question. 
The determination of cell types 
in cell differentiation  processes may be a good candidate. As already
studied by  Kaneko et.al.,\cite{Kaneko} from the dynamical system 
viewpoint, a cell 
society may be modeled by chemical networks with the variable number of 
cells. In these studies, Kaneko et.al. have found that the rule of the 
determination of cell types emerges. We expect that the rule might 
be formalized by state space theory which shares  common features with 
thermodynamics. As developed in our theory concerning thermodynamic
irreversibility,  we hope to find a quantity representing 
a sort of relation associated with biological irreversibility.

\section*{Acknowledgments}

The authors express special  thanks to Y. Oono for thoughtful ideas on 
thermodynamics and enlightening discussions. They acknowledge 
K. Sekimoto, H. Tasaki and S. Takesue for fruitful discussions on
thermodynamic irreversibility;  N. Nakagawa  for helpful
discussions on  dynamical systems; M. Sano and T. Hatano 
for discussions on the fluctuation theorem; T. Chawanya, N. Ito  and K. Sato 
for useful comments on the study; K. Kaneko for stimulating discussions 
on the future perspective;  
P. Gaspard and J. R. Dorfmann for their lectures at Hayama.
One of the authors (T.S.K.) thanks RIKEN for their hospitality
and acknowledges the support from JSPS Research Fellow.

\appendix
\section{}

We summarize  basic properties of the $k$-dimensional volume element
and  $k$-fold exterior product of vectors.\cite{Arnold2}  

Consider a $k$-dimensional surface  in the $n$-dimensional Euclid
space ${\bf R}^n$. The surface can be decomposed into a set of 
sufficiently small pieces of $k$-dimensional parallelaids. 
A $k$-dimensional parallelaid including a point $x \in {\bf R}^n$
is identified to that in a tangent space at  $x \in {\bf R}^n$. 

The $k$-dimensional volume of the projection of the parallelaid 
to a space spanned by $\{ {e}_{i_j},\ 1\le j \le k\}$ is denoted
by $\omega_{i_1i_2\cdots i_k} dx_{i_1}\cdots dx_{i_k} $,
where $dx_{i_1}\cdots dx_{i_k}$ is a $k$-dimensional volume measure
in the space spanned by $\{ {e}_{i_j},\ 1\le j \le k\}$.
A set  $\{\omega_{i_1\cdots i_k} \}$  is called the 
$k$-dimensional volume element. 
 
Let us write the $k$-dimensional volume element in a coordinate-free 
form.  We consider the parallelaid B made by a set of vectors 
$\{b_i, \  1 \le i \le k \}$. We then define the $k$-fold exterior 
product
$b_1 \wedge \cdots \wedge b_k$ as  a map from a $k$-dimensional 
parallelaids to its $k$-dimensional volume of the projection to B. 
Explicitly, for a $k$-dimensional parallelaid  made by a set of vectors 
$\{a_i, \ 1 \le i \le k \}$,  the action of the map 
$b_1 \wedge \cdots \wedge b_k$ is defined as
\begin{equation}
b_1 \wedge \cdots \wedge b_k \cdot (a_1,\cdots a_k)
= {\rm det}{{\cal G}},
\end{equation}
where $G_{ij}=(b_i,a_j)$. 
We write $b_1 \wedge \cdots \wedge b_k$ as $\wedge_{i=1}^k b_i$,
when the order of the vectors is uniquely guessed.   

Then, we can find an exterior product $\wedge_{i=1}^k  \omega_i$  such that 
\begin{equation}
\omega_{i_1\cdots i_k}=
\wedge_{i=1}^k  \omega_i \cdot (e_{i_1},\cdots e_{i_k}).
\end{equation}
Since the exterior product is uniquely determined, 
$k$-dimensional volume element is identified with the $k$-fold 
exterior product. (A set of vectors 
$\{\omega_i, \ 1\le i \le k\} $ is not uniquely determined.)

The $k$-fold exterior product $\wedge_{i=1}^k b_i$ has two
important properties, multi-linearity and skew-symmetry.
The multi-linearity is the relation
\begin{eqnarray}
b_1\wedge\cdots\wedge (c_ib_i+c_i'b_i') \wedge \cdots \wedge b_k
&=&  c_i b_1\wedge\cdots\wedge b_i \wedge \cdots \wedge b_k \\
&+& c_i' b_1\wedge\cdots\wedge b_i' \wedge \cdots \wedge b_k
\end{eqnarray}
for arbitrary $i$, where $c_i$ and $c_i'$ are numbers,
and the skew-symmetry is 
\begin{equation}
\wedge_{l=1}^k  b_{i_l}={\rm sgn}(i_1,\cdots, i_l) 
\wedge_{i=1}^k  b_{i},
\end{equation}
where ${\rm sgn}(i_1,\cdots, i_k) $ takes a value 1 when 
the permutation $(1,\cdots,k) \rightarrow (i_1,\cdots, i_k) $ 
is generated by even number of exchanges, otherwise it takes
a value -1. 

Using the two properties, we can prove 
\begin{equation}
\wedge_{i=1}^k \sum_{j=1}^k G_{ij} b_j
={\rm det}{\cal G} \wedge_{i=1}^k b_i.
\label{pro3}
\end{equation}

(proof)

\begin{eqnarray}
\wedge_{i=1}^k \sum_{j=1}^k G_{ij} b_j &=& 
\sum_{(j_1,\cdots j_k)}
G_{1j_1} {b}_{j_1}\wedge \cdots \wedge G_{kj_k} {b}_{j_k} \\
&=& 
\sum_{(j_1,\cdots j_k)}
G_{1j_1}  \cdots  G_{kj_k}
 {b}_{j_1}\wedge \cdots \wedge  {b}_{j_k} \\
&=&
\sum_{(j_1,\cdots j_k)} {\rm sgn}(j_1\cdots j_k)
G_{1j_1}  \cdots  G_{kj_k} 
 {b}_{1}\wedge \cdots \wedge  {b}_{k} \\
&=&
{\rm det} {\cal G}  {b}_{1}\wedge \cdots \wedge  {b}_{k}.
\end{eqnarray}

\hfill (q.e.d)

The $k$-dimensional volume of the parallelaid B is calculated
as $\sqrt{{\rm det} {\cal B}}$, where $B_{ij}=(b_i,b_j)$. 
We represent  it by $|\wedge_{i=1}^k {b}_i|$. 

(proof)

We can find a set of orthogonal unit vectors 
$\{ u_i, \ 1 \le i \le k \}$ which generate the vector space 
spanned by  $\{ b_i, \ 1 \le i \le k \}$. (One may construct
$\{u_i,\  1 \le i \le k \}$ by employing the Gram-Schmidt procedure.)
Then, the $k$-dimensional volume of B
is given  by
\begin{equation}
|\wedge_{i=1}^k {b}_i|
=|\wedge_{i=1}^k {b}_i \cdot (u_1,\cdots, u_k)|.
\label{defvolB}
\end{equation}
Since $b_i$ can be expanded in the form
\begin{equation}
b_i= \sum_{j=1}^k G_{ij} u_j,
\end{equation}
we obtain
\begin{equation}
\wedge_{i=1}^k {b}_i= {\rm det}{\cal G} \wedge_{i=1}^k u_i,
\end{equation}
where we have used Eq. (\ref{pro3}). Using the identify
\begin{equation}
\wedge_{i=1}^k u_i \cdot (u_1,\cdots, u_k)=1,
\end{equation}
we can rewrite Eq. (\ref{defvolB}) as
\begin{equation}
|\wedge_{i=1}^k {b}_i|
=|{\rm det}{\cal G}|.
\label{defvol2}
\end{equation}
On the other hand, since 
\begin{eqnarray}
(b_i,b_j) &=& \sum_{lm}G_{il}G_{jm}(u_l,u_m), \\
          &=& \sum_{l}G_{il}G_{jl}, \\
          &=& \left( {\cal G}{\cal G}^\dagger \right)_{ij},
\end{eqnarray}         
we have
\begin{equation}
{\cal B}= {\cal G}{\cal G}^\dagger.
\end{equation}
Therefore, the equality
\begin{equation}
{\rm det} {\cal B}= ({\rm det}{\cal G})^2
\end{equation}
holds. Substituting this equality into Eq. (\ref{defvol2}),
we obtain
\begin{equation}
|\wedge_{i=1}^k b_i |=\sqrt{{\rm det} {\cal B}} .
\label{volform}
\end{equation}

\hfill (q.e.d)

Further, for arbitrary $l$ such that   $1 \le l \le k$, 
the inequality
\begin{equation}
| {b}_1\wedge \cdots \wedge  {b}_k|
\le 
| {b}_1\wedge \cdots \wedge  {b}_l|
| {b}_{l+1}\wedge \cdots \wedge  {b}_k|
\end{equation}
holds. This makes us possible to define 'angle' $\phi$ between 
$ {b}_1\wedge \cdots \wedge  {b}_l $ and
$ {b}_{l+1} \wedge \cdots \wedge  {b}_k $ in such a way
that
\begin{equation}
| {b}_1\wedge \cdots \wedge  {b}_k|
=
| {b}_1\wedge \cdots \wedge  {b}_l|
| {b}_{l+1}\wedge \cdots \wedge  {b}_k| \sin \phi.
\end{equation}

(proof)

As  seen in the previous proof, there exist two sets of orthogonal unit
vectors 
$\{  {u}_i, \ 1 \le i \le l\} $  and $\{  {u}_i', \ {l+1} \le i \le k\} $
such that 
\begin{eqnarray}
 {b}_1\wedge \cdots \wedge  {b}_l
&=& | {b}_1\wedge \cdots \wedge  {b}_l|
 {u}_1\wedge \cdots \wedge  {u}_l,  \\
 {b}_{l+1}\wedge \cdots \wedge  {b}_k
&=& | {b}_{l+1}\wedge \cdots \wedge  {b}_k|
 {u}'_{l+1}\wedge \cdots \wedge  {u}'_k,
\end{eqnarray}
where  notice that $ {u}_i $ is not orthogonal to $ {u}'_j$.
Then, we have
\begin{equation}
| {b}_1\wedge \cdots \wedge  {b}_k|
=| {b}_1\wedge \cdots \wedge  {b}_l |
 | {b}_{l+1}\wedge \cdots \wedge  {b}_k|
 | {u}_1\wedge \cdots \wedge  {u}_l \wedge 
 {u}'_{l+1}\wedge \cdots \wedge  {u}'_k|.
\label{app1st}
\end{equation}

Now, by using the Gram-Schmidt orthogonalization,
we define a new set of vectors $\{  {u}_j, \ l+1\le j \le k \}$ as
\begin{equation}
 {u}_j=
{  {u}'_j-\sum_{m=1}^{j-1}( {u}'_j, {u}_m) {u}_m
\over 
s_j
}
\end{equation}
with
\begin{equation}
s_j=|  {u}'_j-\sum_{m=1}^{j-1}( {u}'_j, {u}_m) {u}_m|.
\end{equation}
Here, {}from the equality 
\begin{equation}
|  {u}'_j-\sum_{m=1}^{j-1}( {u}'_j, {u}_m) {u}_m|^2
+|\sum_{m=1}^{j-1}( {u}'_j, {u}_m) {u}_m|^2=1,
\end{equation}
we find
\begin{equation}
0 \le s_j \le 1. 
\label{sj}
\end{equation}
Then, from
\begin{equation}
 {u}'_j=
s_j {u}_j+\sum_{m=1}^{j-1}( {u}'_j, {u}_m) {u}_m,
\end{equation}
we derive
\begin{equation}
 | {u}_1\wedge \cdots \wedge  {u}_k \wedge 
 {u}'_{k+1}\wedge \cdots \wedge  {u}'_k|
=s_{l+1}\cdots s_{k}.
\label{app2nd}
\end{equation}
Substituting Eq. (\ref{app2nd}) into Eq. (\ref{app1st}),
we finally obtain
\begin{eqnarray}
| {b}_1\wedge \cdots \wedge  {b}_k|
&=& | {b}_1\wedge \cdots \wedge  {b}_l |
  | {b}_{l+1}\wedge \cdots \wedge  {b}_k|
  s_{l+1}\cdots s_{k} \\
&\le & | {b}_1\wedge \cdots \wedge  {b}_l |
  | {b}_{l+1}\wedge \cdots \wedge   {b}_k|.
\label{appgoal}
\end{eqnarray}
 
\hfill (q.e.d)



\begin{thebibliography}{99}


\bibitem{Lieb} E. H.  Lieb and J. Yngvason, Phys.Rep. {\bf 310}, 1, (1999).


\bibitem{Lebo} J. L. Lebowitz, Physica {\bf A194}, 1 (1993).


\bibitem{SK} S. Sasa and T.S. Komatsu, Phys. Rev. Lett, {\bf 82}, 912, (1999).

\bibitem{Oono} Y. Oono and M. Paniconi,  Prog. Theor. Phys. Suppl., 
{\bf 130}, 29 (1998).

\bibitem{Ken} K. Sekimoto, { J. Phys. Soc. Jpn.} {\bf 66}, 1234, (1997). 

\bibitem{KS}  K. Sekimoto and S. Sasa,  J. Phys. Soc. Jpn., {\bf 66}, 
 3326, (1997).

\bibitem{SO} K. Sekimoto, Prog. Theor. Phys. Suppl., {\bf 130}, 17, 
(1998);   K. Sekimoto and Y. Oono, (private communication).

\bibitem{Jarz} C. Jarzinski, Phys. Rev. {\bf E56}, 5018, (1997);
Phys. Rev. Lett. {\bf 78} 2690 (1997).

\bibitem{Hatano} T. Hatano, cond-mat/9905012, to appear in 
Phys. Rev. {\bf E }. 

\bibitem{Crooks} G. E. Crooks, Phys. Rev. {\bf E 60 }, 2721, (1999). 

\bibitem{Evans} D. J. Evans, E. G. D. Cohen and G. P. Morriss,
Phys. Rev. Lett., {\bf 71}, 2401, (1993).

\bibitem{Gall} G. Gallvotti and E. G. D. Cohen, J. Stat. Phys., 
{\bf 80}, 931, (1995); Phys. Rev. Lett., {\bf 74} 2694, (1995).

\bibitem{Kurchan} J. Kurchan, J. Phys. {\bf A 31}, 3719, (1998). 

\bibitem{LS} J. L. Lebowitz and H. Spohn, J. Stat. Phys. {\bf 95},
333, (1999). 

\bibitem{Maes} C.  Maes, J. Stat. Phys. {\bf 95}, 367, (1999).

\bibitem{Posch} H.A. Posch and W. G. Hoover, Phys. Rev. {\bf A 38},
473, (1988); {\bf A 39}, 2175, (1989);

\bibitem{ECM} D. J. Evans, E. G. D. Cohen and G. P. Morriss,
Phys. Rev. {\bf A42}, 5990, (1990).

\bibitem{Gas} P. Gaspard, J. Stat. Phys., {\bf 88}, 1215, (1997);
P. Gaspard and G. Nicolis,  Phys. Rev. Lett., {\bf 65} 1693, (1990).

\bibitem{Sato} K. Sato, (private communication).

\bibitem{Sbook} S.Sasa,  {\it Introduction to thermodynamics}, 
(in Japanese) , (Kyoritsu, Tokyo, 2000). (in preparation.)

\bibitem{Tbook} H. Tasaki, {\it Thermodynamics},
(in Japanese), (Baifukan, Tokyo, 2000). (in press.)


\bibitem{LD} Y. Oono, Prog. Theor. Phys. {\bf 99}, 165, (1989),
and references therein.



\bibitem{SN} I. Shimada and T. Nagashima, Prog. Theor. Phys.
{\bf 61}, 1605,  (1979).

\bibitem{GD} P. Gaspard and J. R. Dorfman, Phys. Rev {\bf E52},
3525 (1995). 

\bibitem{Ehren} P. and T. Ehrenfest, {\it The Conceptual Foundation
of the Statistical Approach in Mechanics}, (Dover, New York, 1990);
The original work, in German,  was  published in 1912.

\bibitem{FPU}E. Fermi, J. Pasta and S. Ulam, in {Collected Papers of
Enrico Fermi}, edited by E. Segre (University of Chicago, Chicago, 1965),
vol. 2. p. 978.



\bibitem{Toda}M. Toda, R. Kubo and N. Saito, {\it Statistical Physics I},
(2nd ed.), (Springer-Verlag, Berlin, 1992).

\bibitem{Boltz} L. Boltzmann,  {\it Lectures on Gas theory},
(Dover, New York, 1995); The original work, in German, was
published in 1896 and 1898.

\bibitem{Gibbs} J. W. Gibbs, {\it Elementary principles in statistical
mechanics}, (Dover, New York, 1960); The work was originally published 
in 1902. 


\bibitem{Oseledets} V.I. Oseledets, Trans. Moskow Math. Soc. {\bf 19}, 197,
(1968). 

\bibitem{Pesin} Y. B. Pesin, Russian Math. Surveys {\bf 32:4}, 55, (1977).


\bibitem{demon} H. S. Leff and A. F. Rex eds., {\it Maxwell's Demon;
Information, Entropy, Computing. } (Princeton U.P., Princeton, 1990).




\bibitem{symp} H. Yoshida, Phys.Lett. {\bf A150}, 262, (1990).


\bibitem{Kaneko}K. Kaneko and T. Yomo, J. Theor.Biol., {\bf 199}, 243,
(1999); C. Furusawa and K. Kaneko, Bull.Math.Biology, {\bf 60}, 659, (1998).


\bibitem{Arnold2}V.I. Arnold, {\it Mathematical methods of Classical 
Mechanics},  (New York, Springer-Verlag, 1978).

\end{thebibliography}
\end{document}